\documentclass[a4paper,11pt]{article}

\pdfoutput=1

\usepackage{jheppub}
\usepackage[T1]{fontenc} 
\usepackage{graphicx}
\usepackage{epsfig}
\usepackage{rotating}
\usepackage{amssymb}
\usepackage{subfigure}
\usepackage{dsfont}
\usepackage{psfrag}
\usepackage{amsmath,euscript,array,mathrsfs}
\usepackage{bbold}
\usepackage{epsf}
\usepackage{slashed}

\newcommand{\beq}{\begin{equation}}
\newcommand{\eeq}{\end{equation}}
\newcommand{\beqs}{\begin{eqnarray}}
\newcommand{\eeqs}{\end{eqnarray}}

\newcommand{\lsim}{\mathrel{\raisebox{-
.6ex}{$\stackrel{\textstyle<}{\sim}$}}}
\newcommand{\gsim}{\mathrel{\raisebox{-
.6ex}{$\stackrel{\textstyle>}{\sim}$}}}

\def\di{\mbox{d}}
\def\r{\rho}

\title{Glueballs on the Baryonic Branch of Klebanov-Strassler: dimensional deconstruction and
a light scalar particle}

\author[a]{Daniel Elander,}
\affiliation[a]{Departament de Física Quàntica i Astrofísica \& Institut de Ciències del Cosmos (ICC), \\ Universitat de Barcelona, Martí i Franquès 1, ES-08028, Barcelona, Spain}

\author[b]{Maurizio Piai}
\affiliation[b]{Department of Physics, College of Science, Swansea University, Singleton Park, \\ Swansea SA2 8PP, UK}

\date{\today}

\abstract{
Within gauge/gravity duality, we compute the scalar and tensor mass spectrum in the boundary theory defined by the five-dimensional sigma-model coupled to gravity obtained by constraining to eight scalars the truncation on $T^{1,1}$ that corresponds to the Papadopoulos-Tseytlin (PT) ansatz. We study fluctuations around the 1-parameter family of backgrounds that lift to the baryonic branch of the Klebanov-Strassler (KS) system, and interpolates between the KS background and the Maldacena-Nunez one (CVMN). We adopt a gauge invariant formalism in the treatment of the fluctuations that we interpret as states of the dual theory. The tensor spectrum interpolates between the discrete spectrum of the KS background and the continuum spectrum of the CVMN background, in particular showing the emergence of a finite energy range containing a dense set of states, as expected from dimensional deconstruction. The scalar spectrum shows analogous features, and in addition it contains one state that becomes parametrically light far from the origin along the baryonic branch.
}

\begin{document}
\maketitle
\flushbottom
\allowdisplaybreaks


\section{Introduction}
\label{Sec:introduction}

After the euphoria for the discovery of the Higgs particle~\cite{Higgs},
the LHC program brought us into a new era for particle physics, initiating the exploration of
unprecedentedly high energies. Extensions of the Standard Model (SM) that challenge our current 
understanding of elementary particles and interactions will be put to the test, 
in particular addressing questions about  the fundamental nature of the Higgs itself,
as either an elementary particle or a composite state emerging from a new strongly-coupled theory.
It is hence important to study the mass spectra of controllable strongly-coupled systems with non-QCD-like dynamics, 
to guide our intuition about what to expect on general grounds for realistic models.

An especially interesting open question in the context of quantum field theories at strong coupling
is whether classes of field theories that exhibit large hierarchies between different, dynamically generated, 
physical scales (possibly because of approximate scale invariance,
such as in walking theories~\cite{WTC}), 
 exhibit also the presence of one parametrically light scalar composite state in the mass spectrum, the dilaton.
The identification of the dilaton with the Higgs particle 
would provide a plausible explanation for the current lack of evidence of new physics up to multi-TeV scales,
more than one order of magnitude above the mass of the Higgs particle. 

Gauge/gravity dualities~\cite{AdSCFT,MAGOO} offer a valuable opportunity to
explore non-trivial dynamical features of certain classes of field theories at strong coupling,
by providing a reformulation in terms of  equivalent (dual) weakly-coupled gravity theories in higher dimensional space.
A fully algorithmic procedure allows to 
compute the spectrum of gauge-invariant fluctuations of 
a five-dimensional sigma-model coupled to gravity~\cite{BPM,BHM1,BHM2,E,EP}.
These are interpreted as a subset of the glueballs of the dual field theory.
Their existence is expected on general grounds in large classes of different theories,
and their properties capture important information about the underlying dynamics.
For example, it has been suggested that the ratio of the masses of the lightest spin-0 and spin-2 particles
might highlight the existence of large mass hierarchies in some strongly coupled systems~\cite{ABBELLP}.

Special attention has been devoted in the literature to models in which the supergravity dual is related to the 
conifold and its deformations~\cite{CO,PT,CF,BGGHO,KW,KT,KS,CVMN,HNP,nonsusy}.
Among these, two classes stand out, 
in which a dimension-2 operator~\cite{BGMPZ,MM,GMNP,CNP} 
and/or a dimension-6 operator~\cite{NPP,ENP,NPR,P,EP2,EGNP,E2,A} develop a vacuum expectation value (VEV) (see also~\cite{4cicle}).
We refer to the former as {\it baryonic branch} solutions, for reasons to be explained in the body of the paper,
 and to the latter as {\it walking} solutions, for reasons explained elsewhere~\cite{NPP}, with little bearing 
 in the context of this paper. There exist also explicit constructions of models in which multi-scale dynamics is 
 induced by smearing flavor branes in conifold backgrounds~\cite{R}.

In the case of the dimension-6 operator, evidence that a parametrically light scalar state emerges in the mass spectrum 
has been uncovered in~\cite{ENP,EP2,E2}. Yet, the presence of a mild singularity in the 10-dimensional geometry of such backgrounds obscures its field-theory interpretation.

In this paper, we compute the spectrum of spin-0 and spin-2 states in the boundary theory defined by the five-dimensional sigma-model consisting of eight scalars coupled to gravity obtained from constraining the truncation of the PT ansatz. The limitations of this approach are explained in the body of the paper. We focus on the large class of  backgrounds that lifts to the whole baryonic branch of the KS system~\cite{BGMPZ} --- as well as its extrema corresponding to the CVMN~\cite{CVMN} and KS~\cite{KS} solutions. In the literature of conifold backgrounds, some important features have been discussed for example in~\cite{MM,AD,DKS,GHK}, but the full detailed gravity calculations at strong (field theory) coupling exist only for the KS solution~\cite{K,BHM1,BHM2,spectrumKS} and the 
CVMN solution~\cite{BHM1,BHM2}.

The paper is organized as follows. 
In Section~\ref{Sec:old} and Appendix~\ref{Sec:KSCVMN} we summarize field-theory and supergravity 
results that are known in the literature,
and that allow us to set up the stage for our study. In particular, we fix the notation used later.
In Section~\ref{Sec:asymptotics}, supplemented by Appendix~\ref{Sec:expansions}, \ref{Sec:UV} and \ref{Sec:IR},
 we present in detail the IR and UV expansions of the supergravity backgrounds and of their fluctuations,
that are used in the more technical part of the paper.
In Section~\ref{Sec:spectrum} we present our results for the spectra of spin-0 and spin-2 states,
supplemented in Appendix~\ref{Sec:CVMN} by the study of the scalar spectrum for the CVMN and KS
solutions within the extended sigma-model of this paper.
In Section~\ref{Sec:who} we discuss the field theory interpretation of our results.
We conclude with a discussion of the results and a list of open questions in Section~\ref{Sec:outlook}.

\section{Summary of known results}
\label{Sec:old}

In this section we summarize a set of results that are known from the literature, and 
that we use in the body of the paper, in  the original part of this study.
We refer to the literature for more complete results and discussions.

\subsection{Five-dimensional formalism}
\label{Sec:5}

All the supergravity solutions we consider are described by the Papadopoulos-Tseytlin ansatz (PT)~\cite{PT},
 a subtruncation of the  consistent truncation of type-IIB supergravity on $T^{1,1}$~\cite{CF,BGGHO},
where $T^{1,1}$ is the base of the conifold~\cite{CO}.
The five-dimensional  dynamics
is described by a sigma-model  coupled to gravity, with field content consisting of eight scalar fields $\Phi^a=(\tilde{g},p,x,\phi,a,b,h_1,h_2)$.
The full lift to 10 dimensions can be found elsewhere (see for instance~\cite{EGNP}, from which we borrow the notation, and references therein),
and does not play any role in this paper.

With the conventions of~\cite{BHM1,EP},  the action is
\beqs
\label{eq:5daction}
\int \di^5x\sqrt{-g_5}\left\{
\frac{R}{4}-\frac{1}{2}G_{ab}g^{MN}\partial_M\Phi^a\partial_N\Phi^b-V(\Phi^a)
\right\}\,,
\eeqs
where $R$ is the Ricci scalar, $G_{ab}=G_{ab}(\Phi^a)$ is the sigma-model metric, $g_{MN}=g_{MN}(x^M)$ is the five-dimensional metric,
and where the indexes $a,b=1\,\cdots\,8$ are sigma-model indexes, while $M,N=0,1,2,3,5$ are space-time indexes.
The sigma-model kinetic terms are 
\beqs
\label{eq:sigmamodelGab}
G_{ab}\partial_M\Phi^a\partial_N\Phi^b
&=&
\frac{1}{2}\partial_M \tilde{g} \partial_N \tilde{g} 
\,+\,\partial_M x \partial_N x
\,+\,6\partial_M p \partial_N p\\
&&
\,+\,\frac{1}{4}\partial_M \phi \partial_N \phi 
+\,\frac{1}{2}e^{-2\tilde{g}}\partial_M a \partial_N a+\frac{1}{2}N^2e^{\phi-2x}\partial_M b \partial_N b
\nonumber\\
&&
\nonumber 
+\frac{e^{-\phi-2x}}{e^{2\tilde{g}}+2a^2+e^{-2\tilde{g}}(1-a^2)^2}
\left[\frac{}{}
\frac{1}{2}(e^{2\tilde{g}}+2a^2+e^{-2\tilde{g}}(1+a^2)^2)\partial_M h_2 \partial_N h_2
\right.\\ &&\left.
+(1+2e^{-2\tilde{g}}a^2)\partial_M h_1 \partial_N h_1
+2a(e^{-2\tilde{g}}(a^2+1)+1)\partial_M h_1 \partial_N h_2 \frac{}{} \right]\,,\nonumber
\eeqs
while the potential is
\beqs
\label{eq:sigmamodelV}
V(\Phi^a)&=&-\frac{1}{2}e^{2p-2x}(e^{\tilde{g}}+(1+a^2)e^{-g})
\,+\,\frac{1}{8}e^{-4p-4x}(e^{2\tilde{g}}+(a^2-1)^2e^{-2\tilde{g}}+2a^2)\nonumber\\
&&\,+\,\frac{1}{4}a^2e^{-2\tilde{g}+8p}\,+\,\frac{1}{8}N^2e^{\phi-2x+8p}\left[e^{2\tilde{g}}+e^{-2\tilde{g}}(a^2-2a b +1)^2 +2 (a-b)^2\right]\nonumber\\
&&\,+\,\frac{1}{4}e^{-\phi-2x+8p}h_2^2
\,+\,\frac{1}{8}e^{8p-4x}(M+2N(h_1+b h_2))^2\,.
\eeqs
The free parameters $M$ and $N$ are
related to the fluxes of the $F_5$ and $F_3$ Ramond-Ramond fields respectively~\cite{PT}, and in turn to the size of the gauge groups of the dual theory. Note that $M$ can be eliminated, for $N \neq 0$, by a shift in $h_1$, and $N$ arbitrarily rescaled by appropriate rescalings of $h_1$ and $h_2$, together with a shift of $\phi$. This is an artifact of that we consider only the leading term in the large-$N$ expansion. In the following, we perform this rescaling (shift), so that it effectively puts $M = 0$ and $N = N_c/4 = 1/4$ in Eqs.~\eqref{eq:sigmamodelGab} and \eqref{eq:sigmamodelV}.

The action defined in Eqs.~\eqref{eq:5daction}, \eqref{eq:sigmamodelGab}, and \eqref{eq:sigmamodelV} has been obtained in~\cite{BHM1} by replacing in a more general action a non-linear constraint (see Eq.~(3.11) in~\cite{BHM1}) that effectively removes an extra scalar $\chi$ from the sigma model. The constraint can be derived by varying the action of the consistent truncation on $T^{1,1}$ in~\cite{CF,BGGHO} with respect to one of the vectors, or directly from type-IIB supergravity, but cannot be captured by the dynamics of the sigma-model in five dimensions without extending its field content. This implies that one has to exercise some caution in using the 8-scalar sigma model of Eq.~\eqref{eq:5daction}: it is not necessarily the case that by solving its equations of motion one can automatically construct a solution in type-IIB supergravity. Along the baryonic branch of KS, a complete treatment of the spin-0 spectrum would require to also turn on fluctuations of the additional fields appearing in~\cite{CF,BGGHO}.

In order to find the background solutions of interest, one imposes the general ansatz:
\beqs
\di s^2_5 &=& e^{2A(r)}\di x_{1,3}^2\,+\,\di r^2\,,\\
\Phi^a&=&\bar \Phi^a(r)\,,
\eeqs
in which the background functions depend explicitly on the radial coordinate $r$, but not on the 
four-dimensional  coordinates $x^{\mu}$.
What results is a set of coupled ordinary differential equations, the solution of which determines the background functions
for the metric and the scalar fields in five dimensions.
We will focus on solutions that have an end-of-space at finite $r\rightarrow r_{o}$ and are well defined for all $r>r_o$.

Once the background is fixed, the spectrum of the dual theory is calculated with the following procedure.
As a first step, one allows for small fluctuations of all the scalars $\Phi^a(x^{\mu},r)=\bar \Phi^a(r)+\varphi^a(x^{\mu},r)$,
 as well as the metric by making use of the ADM formalism~\cite{ADM}, 
 and linearizes the resulting second-order equations for the fluctuating fields.
These are conveniently rewritten in terms of  gauge-invariant combinations following~\cite{BHM1,E}.
Having decomposed the five-dimensional metric in its four-dimensional components, 
and Fourier-tranformed in the four Minkowski directions,
 the transverse and traceless part $\mathfrak{e}^{\mu}_{\,\,\,\,\nu}$  of the metric fluctuations
 must obey
\beqs
\left[\frac{}{}\partial_r^2+4\partial_rA\partial_r+e^{-2A}m^2\right]\mathfrak{e}^{\mu}_{\,\,\,\,\nu}&=&0\,,
\eeqs
where $m^2=-q^{\mu}q^{\nu}\eta_{\mu\nu}$, in terms of the four-dimensional momentum $q^{\mu}$.
The dynamical components of the  gauge-invariant combinations of the scalars, resulting from the mixing between 
the sigma-model fluctuations $\varphi^a$ and the scalar component $h$ of the metric, are denoted~\cite{BHM1,E}
\beqs 
\mathfrak{a}^a(q^{\mu},r)\equiv\varphi^a(q^{\mu},r)-\frac{\partial_r\bar{\Phi}^{a}}{6\partial_rA^{}} h(q^{\mu},r)\,,
\label{Eq:a}
\eeqs
with $a$ the sigma-model index,
and they obey:
\beqs
\label{Eq:diffeq}
	0&=&\Big[ {\cal D}_r^2 + 4 \partial_{r}A {\cal D}_r + e^{-2A} m^2 \Big] \mathfrak{a}^a \\ \nonumber
	&& - \Big[ V^a_{\ |c} - \mathcal{R}^a_{\ bcd} \partial_{r}\bar\Phi^b \partial_{r}\bar\Phi^d + \frac{4 (\partial_{r}\bar\Phi^a V^b + V^a 
	\partial_{r}\bar\Phi^b) G_{bc}}{3 \partial_{r} A} + \frac{16 V \partial_{r}\bar\Phi^a \partial_{r}\bar\Phi^b G_{bc}}{9 (\partial_{r}A)^2} \Big] \mathfrak{a}^c\,.
\eeqs
The notation is explained in detail in~\cite{EP}: ${\cal D}_r$ is the background covariant derivative, 
$V_a\equiv\partial V/\partial \Phi^a$, ${\cal R}^a_{\ bcd}$ is the Riemann 
curvature tensor with respect to the sigma-model metric, and $V^a_{\,\,\,|b}\equiv 
D_b(G^{ac} V_c)$, with $D_b$ the sigma-model covariant derivative.

As a second step, one introduces two unphysical cutoff scales $r_{I,U}$, 
by assuming the radial direction $r$  of the geometry be bounded as in $r_o<r_I<r< r_U<\infty$.
One imposes the boundary conditions for the tensors ($i = I, U$)
\beqs
\left.\frac{}{}\partial_r\mathfrak{e}^{\mu}_{\,\,\,\,\nu}\right|_{r=r_i}&=&0\,,
\eeqs
and for the scalars~\cite{EP}
\beqs
\label{Eq:BCb}
 \frac{2  e^{2A}\partial_{r}\bar \Phi^a }{3m^2 \partial_{r}A}
	\left[ \partial_{r}\bar \Phi^b{\cal D}_r -\frac{4 V \partial_{r}\bar \Phi^b}{3 \partial_rA} - V^b \right] \mathfrak a_b +\mathfrak a^a\Big|_{r_i} = 0 \, .
\eeqs
These boundary conditions were derived in~\cite{EP} by requiring that the variational problem be well defined. More precisely, one has to add to the five-dimensional action two boundary-localized four-dimensional contributions, the structure of which is fixed by consistency requirements up to a choice of quadratic terms for the sigma-model fields. Taking these boundary-localized mass terms to infinity leads to the boundary conditions $\varphi^a |_{r_i} = 0$, which rewritten in terms of the gauge invariant variable $\mathfrak a^a$ is equivalent to Eq.~\eqref{Eq:BCb}. As one might expect, adding infinite mass terms localized at the boundaries assures that the least divergent modes of the fluctuations are selected. Indeed, we will show explicitly in Section~\ref{Sec:asymptotics} that this procedure is equivalent, in the present context and after taking $r_I\rightarrow r_{o}$ and $r_{U}\rightarrow +\infty$, to requiring regularity and normalisability, according to the standard prescription of gauge/gravity dualities. It also ensures the absence of accidental (fine-tuned) cancellations in the calculation of the mass spectrum (see discussion in Section 5.2 of~\cite{EFHMP}).

The cutoffs have no physical meaning, but are necessary as regulators, for two technical reasons:  the backgrounds of interest  can be found only by solving the differential equations numerically, and furthermore the five-dimensional solutions
 of interest  (but not the 10-dimensional lifts) are singular both in the IR (small $r$) and in the UV (large $r$), and it is hence necessary to perform the calculations with finite regulators.

As a final step, one studies numerically the spectrum by scanning over $m^2$ in discrete steps. 
For each value of $m^2$ one independently evolves the solutions of the bulk equations having imposed the IR and UV 
boundary conditions, and one tests whether the two match (including their first derivatives) at some intermediate value of $r$.
Importantly, one must repeat the calculations by varying the UV cutoff $r_U$ (and IR cutoff $r_I$) towards larger (smaller) values.
This process is morally equivalent to the standard study of finite spacing and finite volume systematic effects
in lattice gauge theories. 

If the extrapolation of the numerical output from finite cutoff can be done, so that what results is a spectrum that is
independent of the position of the (unphysical) cutoffs, then this is interpreted as the spectrum of tensor and scalar bound states in the dual field theory. Otherwise, one has to conclude that the physical cutoffs 
 must be kept in place and are an essential part of the dynamics of the full theory. This may happen because of a bad singularity
 in the IR of the geometry, such as is the case for the GPPZ model~\cite{GPPZ,PW}.\footnote{While the calculation of the spectrum in~\cite{MP} for the 1-scalar truncated system
 is convergent, see for instance~\cite{EP3} for a calculation of the spectrum
 of the truncation to two scalars with non-trivial bulk profile, in the limit in which the mass deformation is large.
 In this case, the spectrum contains a state the mass of which depends on the IR cutoff in such a way that the cutoff cannot be removed.}
 Similar problems emerge in backgrounds that are badly behaved in the 
 UV (see for instance in~\cite{EP2}  the discussion of the spectrum of wrapped-D5 backgrounds 
 with constant-dilaton asymptotic behavior). Unremovable cutoff dependences emerge also in some lattice gauge theories,
 for example in the presence of bulk phase transitions. In all  these cases, one cannot  
 trust the calculations to represent the dynamics of a continuum four-dimensional field theory.
 We anticipate here the fact that a smooth behavior appears 
 for the backgrounds discussed in the present paper,
so that both regulators can be removed, and the results we will present later admit a trustable field-theory interpretation.

In the specific case at hand, it is convenient to perform the change of variable
\beqs
\di r &=& 2 e^{-4p} \di \r\,,
\eeqs
hence rewriting of the equations for the  tensorial fluctuations as
\beqs
\left[\frac{}{}\partial_{\rho}^2+ \left[4\partial_{\r}A - \partial_{\r} \log(\partial_{\r} r) \right] \partial_{\rho}+ (\partial_{\r} r)^2 e^{-2A}m^2\right]\mathfrak{e}^{\mu}_{\,\,\,\,\nu}&=&0\,.
\eeqs


For practical purposes, we also  rewrite the bulk equations 
for the scalars as~\cite{E}:
\beqs
\left[\frac{}{}\delta^a_{\,\,\,\,b}\partial_{\r}^2+S^a_{\,\,\,\,b}\partial_{\r}+T^a_{\,\,\,\,b}+ (\partial_{\r} r)^2 e^{-2A}m^2 \delta^a_{\,\,\,\,b} \right] \mathfrak{a}^b\,=\,0\,,
\label{Eq:ST}
\eeqs
where the matrices $S^a_{\,\,\,\,b}$ and $T^a_{\,\,\,\,b}$ are defined by
\beqs
\label{eq:SandT}
S^a_{\,\,\,\,b}&=&2{\cal G}^a_{\,\,\,\,bc}\partial_{\rho}\bar{\Phi}^c\,+\, \left[4\partial_{\r}A - \partial_{\r} \log(\partial_{\r} r) \right] \delta^a_{\,\,\,\,b}\,,\\
T^{a}_{\,\,\,\,b}&=&\partial_b{\cal G}^a_{\,\,\,\,cd}\partial_{\r}\bar{\Phi}^c\partial_{\r}\bar{\Phi}^d
- (\partial_{\r} r)^2 \left[\frac{}{}\left(\frac{4(V^a\partial_{\r}\bar{\Phi}^c+V^c\partial_{\r}\bar{\Phi}^a)}{3\partial_{\rho}A}+
\frac{16V\partial_{\r}\bar{\Phi}^a\partial_{\r}\bar{\Phi}^c}{9(\partial_{\r}A)^2}\right)G_{cb}\,+\partial_b V^a\right]\,.\nonumber
\eeqs


From now on, we adopt the convention that $F^{\prime}\equiv \partial F/\partial \r$ for any  $F$,
and we write all background functions and fluctuations in the variable $\r$.

\subsection{Background solutions}
\label{Sec:bg}

The general form of the solutions of interest can be written via an algebraic process involving two functions of the radial coordinates
$P(\r)$ and $Q(\r)$, a few hyperbolic functions, and  the integration constants $\kappa_2$, $\phi_o$ and $A_0$.
We report them here explicitly:
\beqs
	\tilde{g}(\rho) &=& \frac{1}{2} \log \left(\frac{P(\rho )^2-Q(\rho )^2}{(P(\rho )
   \coth (2 \rho) - Q(\rho))^2}\right) , \\
   	p(\rho) &=& -\frac{\phi_o}{6} - \frac{1}{24} \log \left(\frac{\hat{h}(\rho)^4 P'(\rho )^3 \left(P(\rho )^2-Q(\rho )^2\right) \sinh^2(2 \rho)}{131072 }\right) , \\
   x(\rho) &=& \frac{\phi _o}{2} + \frac{1}{8} \log \left(\frac{\hat{h}(\rho )^4 \sinh ^2(2 \rho )
   \left(P(\rho )^2-Q(\rho )^2\right)^3}{8192 P'(\rho
   )}\right), \\
   \phi(\rho) &=& \phi _o-\frac{1}{4} \log \left(\frac{P'(\rho
   ) \left(P(\rho )^2-Q(\rho )^2\right)}{8\sinh^2(2 \rho)} \right) , \\
   a(\rho) &=& \frac{P(\rho ) \rm{csch}(2 \rho )}{P(\rho ) \coth (2 \rho )-Q(\rho )} , \\
   b(\rho) &=& \frac{2 \rho}{ \sinh(2 \rho)} , \\
   h_1(\rho) &=& \frac{\kappa_2 e^{2 \phi _o} \cosh (2 \rho ) Q(\rho )}{\sqrt{2}
   \sqrt{P'(\rho ) \left(P(\rho )^2-Q(\rho )^2\right)}} , \\
   h_2(\rho) &=& -\frac{\kappa _2 e^{2 \phi _o} Q(\rho )}{\sqrt{2} \sqrt{P'(\rho )
   \left(P(\rho )^2-Q(\rho )^2\right)}} , \\
   A(\rho) &=& A_0 + \frac{2\phi_o}{3} +\frac{1}{6} \log \left(\frac{1}{16} \hat{h}(\rho ) \sinh ^2(2
   \rho ) \left(P(\rho )^2-Q(\rho )^2\right)\right) ,
\eeqs
where $\hat{h}(\rho) = 1- \kappa_2^2 e^{2\phi(\rho)}$. Here, we have chosen an integration constant in such a way that the space ends at $\rho = 0$. Furthermore, integration constant $A_0$ has no physical meaning, as it can be reabsorbed into a rescaling of the field theory coordinates $x^{\mu}$.

The functions $Q$ and $P$ are obtained by solving the first-order (BPS) equations of type-IIB supergravity~\cite{HNP}. The function $Q$ is given by
\beqs
Q(\rho) &=& 2\r\coth(2\r)-1 \,,
\eeqs
where we have fixed an integration constant so as to avoid a bad IR singularity, while the function $P$ obeys the following non-linear second-order {\it master}
equation~\cite{HNP}:
\beqs
P^{\prime\prime}+P^{\prime}\left(\frac{P^{\prime}+Q^{\prime}}{P^{\prime}-Q^{\prime}}+
\frac{P^{\prime}-Q^{\prime}}{P^{\prime}+Q^{\prime}}-4\coth(2\r)\right)\,=\,0\,,
\label{Eq:master}
\eeqs
the generic solution of which depends on two additional integration constants.

\begin{figure}[t]
\begin{center}
\includegraphics[height=6.8cm]{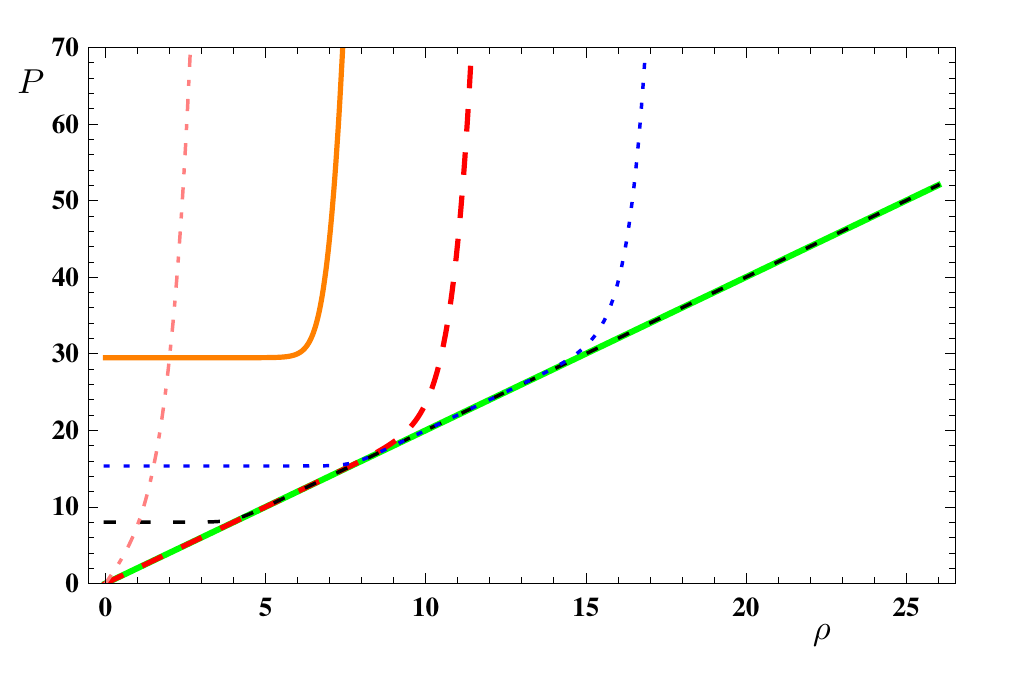}
\caption{Six examples of solutions to the master equation representative of
all the classes discussed in Section~\ref{Sec:old}.
In green the function $P=2\r$, the CVMN solution. In black (short dashing) a solution that 
has the CVMN behavior for $\r>\r_{\ast}\simeq 4$, but has $P\sim P_0$ for $\r<\r_{\ast}$.
In red (long dashing) a solution that has the CVMN behavior for $\r<\bar{\r}\simeq 9$, but behaves as $P\propto e^{4\r/3}$ for $\r>\bar{\r}$.
In  blue (dotted) a solution that has $P\simeq P_0$ for $\r<\r_{\ast}\simeq 8$, behaves as $P\simeq 2\r$ for $\r_{\ast}<\r<\bar{\r}\simeq 14$,
and then behaves as $P\propto e^{4\r/3}$ for $\r>\bar{\r}$.
In orange a solution for which $P\gg 2\r$ for all $\r$, such that $P\simeq P_0$ for $\r<\r_{\ast}\simeq 6$ 
and $P\propto e^{4\r/3}$ for $\r>\r_{\ast}$. In pink (dot-dashed) a solution that has $P\propto e^{4\r/3}$ obtained with $\kappa_1=6$ in Eq.~(\ref{Eq:baryonic}).}
\label{Fig:illustrative}
\end{center}
\end{figure}

We find it useful to remind the reader about what is known for all 
interesting solutions $P$. We focus on solutions for which $P$ and $P^{\prime}$ are both smooth and monotonically non-decreasing for all $\r\geq 0$.
Locally (for $\rho \gsim 1$) the only acceptable solutions are given by the following three possibilities (for a more general 
and precise discussion, see for instance~\cite{FPS}):
\begin{itemize}

\item[a)] constant $P\simeq P_0$,

\item[b)] linear $P\simeq 2\r$,

\item[c)] exponential $P\sim e^{4\r/3}$.

\end{itemize}
Furthermore, for all $\r>0$ one finds that it is necessary to impose the constraint $P(\r)\geq 2\r$.
Various examples of admissible solutions of the master equation,
illustrating all possible qualitative behaviors, are shown in Fig.~\ref{Fig:illustrative}.

In this paper, we  perform most of our calculations by making use of solutions $P$ that 
asymptotically grow exponentially for large $\r$. We hence report here the UV expansion of such solutions:
\beqs
P&=&3 c_+ e^{4\r/3}\,+\,\frac{4}{3c_+}\left(\r^2-\r+\frac{13}{16}\right)e^{-4\r/3}\,-\,\left(8c_+\r + \frac{c_-}{192c_+^2}\right)e^{-8\r/3}\nonumber\\
&&
\,+\,\frac{1}{c_+^3}\left(\frac{2063}{1536}+\frac{103}{32}\r+\frac{1}{4}\r^2+\frac{4}{3}\r^2\right)e^{-4\r}\nonumber\\
&&\,+\,e^{-16 \r/3} \left(+\frac{5815 c_--435456 c_+^3}{1728000 c_+^4}
   +\frac{\r \left(1177344 c_+^3-1410 c_-\right)}{259200
   c_+^4}
   \nonumber\right.\\
   &&\left.
   +\frac{\r^2 \left(45 c_--152064 c_+^3\right)}{19440
   c_+^4}
   +\frac{32
   \r^3}{9 c_+}\right) \label{Eq:P}\\
&&\,+\,e^{-20 \r/3} \left(\frac{1}{32} \left(-\frac{27 c_-^2+2698210}{93312  c_+^5}-32 c_+\right)
+\frac{1}{48} \r \left(-\frac{4 c_-}{3  c_+^2} -\frac{7457}{324 c_+^5}\right)
\nonumber\right.\\
   &&\left.
+\frac{1}{72} \left(\frac{60305}{432
   c_+^5}-1536 c_+\right) \r^2   -\frac{7495 \r^3}{5832 c_+^5}
   +\frac{145 \r^4}{243  c_+^5}
    -\frac{80 \r^5}{243 c_+^5}
 - \frac{64 \r^6}{729 c_+^5}
 \right)
 \,+\,{\cal O}(e^{-8\r})\,,\nonumber
\eeqs
where $c_+$ and $c_-$ are two integration constants.
We will restrict our attention to backgrounds that are completely smooth (in 10 dimensions),
which requires forbidding the behavior a) for $P(\r)$. In turn, this means that the constant $c_-$ is fixed by the requirement of 
regularity in the IR, and only $c_+$ is free.

We observe  that $\phi_o \rightarrow \phi_o + \delta\phi_o$ corresponds to a shift under which the 5-dimensional
 action remains the same up to an overall multiplicative factor that could be reabsorbed into the definition of  the 5-dimensional Planck scale.
Therefore, without loss of generality, we fix the asymptotic value of the 10-dimensional dilaton
 to be $\phi_\infty = 0$ for all backgrounds we consider, except for the CVMN one (for which $\phi$ grows linearly in the UV). By using the large-$\r$ behavior of $P$ from Eq.~(\ref{Eq:P}), it follows that we fix
\beq
	\phi_o \equiv \frac{1}{4} \left( \log(18) + 3 \log c_+ \right) .
\eeq
With all of this in place, the range of admissible values for $\kappa_2$ is now $0\leq \kappa_2 \leq 1$ in order to ensure that the background scalars be real.
We fix $\kappa_2=1$ in the following, unless explicitly stated otherwise, and in this way the UV asymptotics of the background functions reproduces the KS ones.

We consider solutions of the master equation~(\ref{Eq:master}) 
that in the IR take the form~\cite{HNP}:
\beqs
P&=&\kappa_1\r\,+\,\frac{4\kappa_1}{15}\left(1-\frac{4}{\kappa_1^2}\right)\r^3\,+\cdots\,,
\label{Eq:baryonic}
\eeqs
where $\kappa_1\geq 2$ is an integration constant.
By varying $\kappa_1$ one obtains for example the solutions in pink, red and green in Fig.~\ref{Fig:illustrative}.
For $\kappa_1=2$  one  recovers the CVMN solution. As long as $\kappa_1$ is close to $2$, there exists a range $0<\r\lsim \bar{\r}$ along which $P$ is approximately  linear, before the exponential behavior takes over at large $\r>\bar{\r}$.  Approximately, we find that $\bar{\r}\simeq 1-\frac{1}{2}\log(\kappa_1-2)$. The baryonic branch solutions in~\cite{BGMPZ} are obtained by choosing $\kappa_1>2$ (and setting $\kappa_2=1$). Conversely, the KS solution can be reproduced by a limiting procedure that is equivalent to taking ${\kappa}_1 \rightarrow +\infty$.

\begin{figure*}[h]
\begin{center}
\includegraphics[height=18.1cm]{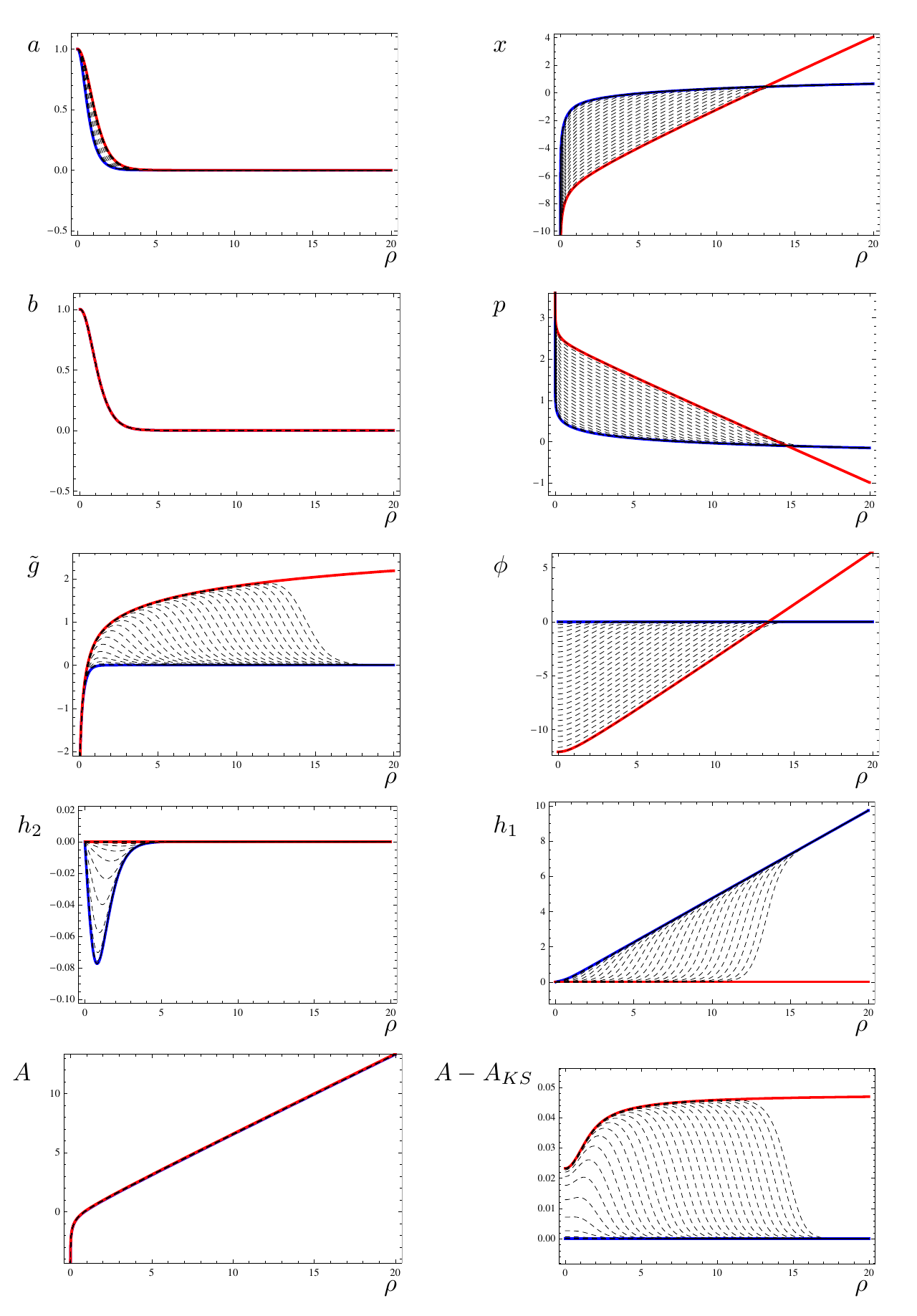}
\caption{All the background functions for several examples of backgrounds:
in blue we exhibit the KS solution, in black (dashing) several examples of baryonic branch solutions with
different values of $\kappa_1$ (and hence $\bar{\r}$), and in red the CVMN solution.
See the main text for clarifications about the choices of integration constants adopted.
Notice that in the bottom left panel all curves for $A$ are on top of each other. To make visible the
small differences, we show in the bottom-right panel the difference $A-A_{KS}$, on a much smaller scale.
 }
\label{Fig:baryonic}
\end{center}
\end{figure*}

For illustration purposes we  show some comparisons 
of the CVMN solutions --- for which the dilaton grows linearly in the UV --- to the baryonic branch and KS ones in Fig.~\ref{Fig:baryonic}.
To do so, we adjust $\phi_o$ for the CVMN solution to match
the background that has the smallest $c_+$ (largest $\bar{\rho}$).
Only the appearance of $x$, $p$ and $\phi$ is (somewhat artificially) affected, 
facilitating the comparison of the CVMN 
solution to the baryonic branch one.

\subsection{The baryonic branch}
\label{Sec:baryonic}

\begin{figure}[t]
\begin{center}
\includegraphics[height=7.2cm]{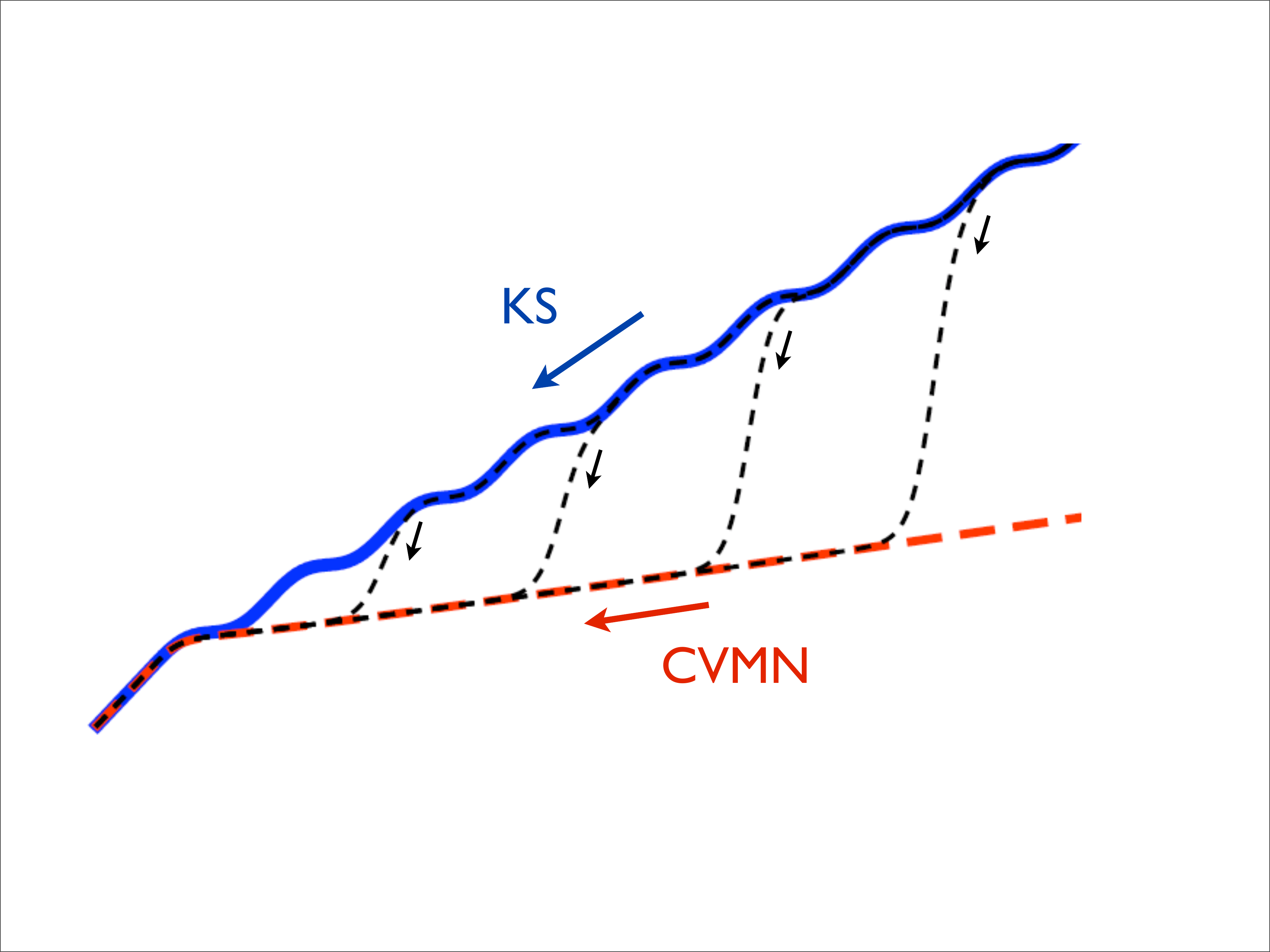}
\caption{Illustrative depiction of the RG flows of the field theories discussed in the paper:
the KS solution (blue, continuous line), the CVMN solution (red, long-dashed line), and a few examples of the baryonic branch solutions for different values of $q$ (black, short-dashed line).}
\label{Fig:figBARYONIC}
\end{center}
\end{figure}

We summarize  in this section some known field-theory and supergravity notions about
the baryonic branch solutions that are important to understand and interpret our results in the subsequent sections of the paper.

The four-dimensional  field theory dual to the supergravity solutions in the PT ansatz is described for example in~\cite{DKS}. 
It consists of a ${\cal N}=1$ supersymmetric quiver gauge theory with two gauge groups and a set of bifundamental 
matter fields realizing a global $SU(2)\times SU(2)$ symmetry.
The gauge group is $SU(k N)\times SU((k+1)N)$, with $N$ and $k$ positive integers.
The running of the gauge couplings towards the IR 
undergoes the {\it duality cascade}~\cite{KS,S},
namely the dynamics can be described in terms of a chain of effective field theories
(Seiberg dualities) with sequentially smaller gauge group 
$SU(k N)\times SU((k+1)N)\rightarrow SU(k N)\times SU((k-1)N) \rightarrow SU((k-2) N)\times SU((k-1)N) \rightarrow \cdots$.
Ultimately, one reaches the stage of the cascade at which the gauge group is $SU(2N)\times SU(N)$,
at which point one can discuss the meson and baryon operators and their effective superpotential,
which takes into account  the dynamically generated scale $\Lambda$~\cite{DKS}.
 The theory confines and the (dimension-3) gaugino condensate forms. The vacuum structure is non-trivial,
 as several inequivalent vacua are allowed.

To better understand the vacuum structure requires looking carefully at the symmetries of the system.
Besides the aforementioned ones, there is a $U(1)_B$ corresponding to baryon number,
which is exact, and spontaneously broken by the VEVs of the baryon operators.
There is a $U(1)_R$, anomalously broken to $Z_{2N}$ and spontaneously broken to $Z_2$ by the gaugino condensate.
And finally there is an additional exact discrete $Z_2$ that is related to the exchange of the two global $SU(2)$ symmetry groups,
and that characterizes the KS limit of the baryonic branch. 

The moduli space of the theory contains, in particular, a baryonic branch: the operator $\cal U$ defined
by Eq.~(4.2) in~\cite{DKS} may develop a VEV~\cite{BGMPZ,MM,DKS}, causing an imbalance in the 
two baryon condensates and breaking the $Z_2$ symmetry of the KS system.
The $U(1)_B$ baryon symmetry is exact and spontaneously broken along the whole baryonic branch, but $\cal U$ also triggers the higgsing of the gauge symmetry $SU(q N)\times SU((q+1)N)\rightarrow SU(N)$ for some value of $k=q$. At this point, the infinite chain of Seiberg dualities stops, and the gauge bosons in the coset acquire a mass and decouple from the dynamics. The scale of the condensate $\cal U$ is controlled not only by the dynamical scale $\Lambda$ of the theory but also the parameter $q$ along the infinite moduli space. Furthermore, at the perturbative level, it can be shown that the non-vanishing fields carrying baryon number assume VEVs that reproduce the algebra of $SU(2)$~\cite{DKS,MM}. For finite $q$
the spectrum of massive modes due to higgsing on the baryonic branch deconstructs a 2-sphere,
in the sense that subsets of the massive modes span a finite set of finite-dimensional representations of $SU(2)$
(see also the  explicit  calculations  in~\cite{MM}, and the analogy with~\cite{AD}).

The general picture one expects in QFT terms, presented in the cartoon in Fig.~\ref{Fig:figBARYONIC},
 is hence of a theory that admits an infinite number of possible dynamical
realizations, because of the 1-parameter moduli space. In the far-UV, all of them are undergoing an
infinite number of Seiberg dualities, until at some special value of $k=q$ the Higgsing makes the cascade stop.
In the dual theory, this is represented by the departure of the background from resembling the KS solution ($P\propto e^{4/3\r}$,
$\kappa_2=1$),
and rather resembling the CVMN one ($P\sim 2\r$).
In the deep IR the theory confines, and the gaugino condensate appears at scale $\Lambda$.
When $q=0$, the dual gravity picture is provided by
the KS solution. In the  limit in which $q\rightarrow+\infty$ the dual is described by the CVMN
solution~\cite{CVMN}: in this case the representations of $SU(2)$ are infinite-dimensional, and the dual gravity
description is obtained by wrapping a stack of $D5$ branes on a 2-sphere in the internal geometry 
and then taking the strong-coupling limit.
The baryonic branch solutions in~\cite{BGMPZ} interpolate between the KS and CVMN solutions, for finite $q$.

This dynamical behavior at finite points $q$ along the baryonic branch results in a low-energy description of the
theory characterized by two distinct, parametrically separated, physical scales: besides the scale $\Lambda$,
controlling the mass gap in the theory, a second scale controlled by $q$ appears, related to the mass of the heaviest 
mode coming from the deconstruction of the 2-sphere. 
In practical terms, this mechanism results in a natural way to produce a large hierarchy of scales, in which the 
 number of KK modes included in the deconstruction of the 
sphere plays the role of a tunable parameter $q$.  

One expects the mass spectrum of individual modes to be characterized by three distinct energy intervals.
At low energies below $\Lambda$, the 
best description of the system is in terms of a four-dimensional effective field theory containing 
only a few discrete, light bound states.
At intermediate energies, over a range controlled by $q$, one expects the densely-packed spectrum 
of bound states resulting from the deconstruction of a six-dimensional theory on a 2-sphere. 
At further higher energies one expects to recover the 
typical structure of bound states of the KS case, the supergravity dual of a four-dimensional theory. 
Notice that the latter is not the Regge behavior,
as the masses of the bound states will scale as $M^2_j\propto j^2$, where $j$ is the excitation number, as opposed to
$M^2_j\propto j$ as is expected in string theory: what one is computing here are the supergravity excitations of the 
lightest stringy modes  that are retained in supergravity, 
and hence only a subset of the glueballs are captured.

In supergravity, the baryonic branch is characterized by the non-vanishing of the 
background field dual to the  dimension-2 operator ${\cal U}$, that is represented by the combination
\beqs
v_2&\equiv&a^2+e^{2\tilde{g}}-1\,=\,\frac{2Q}{P\coth(2\r)-Q}\,.
\eeqs
In order to recover the KS system, one has to set $v_2=0$.
The vanishing of $v_2$ is related to the very existence of a further subtruncation of the five-dimensional PT
system to the KS one, that admits only seven scalar fields: the resulting constraint 
amounts to the presence of a $Z_2$ symmetry (as anticipated)
that in the underlying 10-dimensional geometry relates to the exchange of the two $S^2$ within $T^{1,1}$.
This symmetry is broken along the baryonic branch, a fact that will play a crucial role later in the paper (see also~\cite{GHK} for useful discussions
on this point).

It is now time to the make transparent the meaning of $\kappa_2$, starting by noticing that $v_2$ does not depend upon it.
This part of the discussion draws heavily on the arguments outlined in~\cite{MM,GMNP,EGNP}.
From the expression of $v_2$ one sees that along the solutions in Eq.~(\ref{Eq:baryonic}) 
for $\r<\bar{\r}$ it is unsuppressed, but the exponential growth of $P$
for $\r>\bar{\r}$ forces it to switch off at large scales.

For generic values of  $0\leq \kappa_2<1$,
the asymptotic behavior in the UV of the supergravity solutions
does not admit a simple field theory interpretation, and yields to the kind of pathologies one expects by extrapolating to high energies
the behavior of a (UV-incomplete, dual) effective field theory. 
For example neither the probe-string prescription for computing the expectation value of
the Wilson loop in the dual field theory~\cite{Wilson}, nor the calculation of the spectrum of fluctuations of the background
to obtain the glueballs can be perfomed, unless one has a finite, physical cutoff in the far UV of the theory~\cite{EP2}.
Indeed, the analysis of the operators near the Klebanov-Witten fixed point~\cite{dimension} shows that 
in these solutions there is a dimension-8 operator, the coupling of which is non-trivial, and that makes the field
theory UV-incomplete.
The dynamical origin of this fact resides in the observation that in the presence of a VEV for $\cal U$ one can integrate out 
massive degrees of freedom: the wrapped-$D5$ solutions  with this asymptotic behavior provide the dual of the resulting EFT.
By dialing $\kappa_2=1$ one  removes the dimension-8 operator from the field theory, and replaces it 
by reinstating the  heavy gauge bosons of the quiver.
The baryonic branch solutions in~\cite{BGMPZ} are hence obtained by looking at solutions for $P$ of the form of
Eq.~(\ref{Eq:baryonic}), and setting $\kappa_2=1$ in such a way as to smoothen the UV behavior of the theory
to reproduce the duality cascade for $\r > \bar \r$~\cite{GMNP,EGNP}.

The first purpose of this paper is to compute the mass spectrum of  tensor modes on the baryonic branch.
We will explicitly check that in the two appropriate limits our results reproduce the known ones for the KS and CVMN case.
Namely, in the KS case it is known that the spectrum is discrete, and has been studied in detail~\cite{K,BHM2,spectrumKS}, 
while in the CVMN case
the spectrum has no discrete (bound) states, but exhibits a mass gap, beyond which a continuum appears~\cite{BHM1}.
The latter is  a manifestation of the fact that at energies far above the confinement scale
the field theory described by the CVMN system is six dimensional, as is apparent in gravity from
the fact that the internal $S^2$ is blowing up towards the UV.
While it is known that the spectrum computed perturbatively reproduces the features of a sphere~\cite{AD}, we want to show
that this holds true also non-perturbatively. In particular, we expect the spectrum of the generic baryonic branch solution
to exhibit a mass gap at low energies, followed by a region with high density of states at intermediate energy, 
followed again at  high energies by the discrete spectrum of bound states typical of the supergravity dual description of a confined gauge theory in four dimensions.

The spectrum of scalar states on a generic point on the baryonic branch has not been computed so far in the literature, and as of now it remains an open problem. Our second purpose is to perform this calculation (with the caveats discussed earlier) in the 8-scalar sigma model corresponding to the constrained PT ansatz, and to see to which extent such calculation captures the qualitative features expected from field theory arguments. In particular, we want to check that the spectrum interpolates between the known KS and CVMN cases, we want to assess whether we see evidence of the emergence of dimensional deconstruction and, last but not least, we want to understand whether the position along the moduli space of dual field theory is associated with an anomalously light scalar state that can be interpreted as a pseudo-dilaton.

\section{Asymptotic expansions}
\label{Sec:asymptotics}

The nature of the differential equations controlling the system
is such that it is not practically possible to solve them numerically at arbitrarily large (small) values of the 
radial direction $\r$. In order to address this technical limitation, we  supplement our  treatment 
by making use of the asymptotic expansions of both background and fluctuating fields that enter the dynamical system,
in a process that is reminiscent of what in the lattice literature is commonly referred to as {\it improvement}~\cite{improvement}.
For this purpose, in this section and in the related Appendix~\ref{Sec:expansions}, \ref{Sec:UV} and \ref{Sec:IR},
we provide the reader with the explicit form of the  expansions for the relevant 
quantities and discuss their salient features.

We start with the background functions.
In the UV, we find it convenient to introduce
a new radial coordinate $z\equiv e^{-\frac{2}{3}\r}$. 
We  list in Appendix~\ref{Sec:expansions} the explicit expansion of the solutions at large $\r$.
Several things are worth highlighting.
First of all, the warp factor does not behave as in asymptotically AdS space with $A\sim \frac{1}{z}$,
but exhibits a logarithmic correction. This is the effect of the duality cascade: the RG flow towards the UV
follows closely a line of fixed points describing Klebanov-Witten CFTs~\cite{KW}, but strictly speaking the theory is not 
UV complete, as the imbalance between the two gauge groups cannot be removed (see~\cite{S} for a pedagogical and clear explanation).

Yet, one sees that the background is close enough to  AdS  that one can infer the dimensionality of several operators by
looking at the expansions of the corresponding scalars in the background.
We recall that the studies in~\cite{dimension} conclude 
 that $\tilde{g}$ is associated with a dimension-2 operator, $\phi$ and $h_1$ with two dimension-4 operators,
$a$ with a dimension-3 operator, $b$ and $h_2$ are the result of mixing between a dimension-3 and a dimension-7 operator,
and finally $x$ and $p$ are related to  mixing between a dimension-6 and a dimension-8 operator. 
Notice the appearance of $\log(z)$ terms in the expansion.


By expanding the backgrounds on the baryonic branch (with $\kappa_2=1$)
near the end-of-space in the IR, one finds the expressions in Appendix~\ref{Sec:expansions}.
Several of them are singular, yet, by comparing with the expression for the metric 
in the 10-dimensional language (see for instance~\cite{EGNP}), one can be convinced that the 10-dimensional metric is smooth.


\subsection{Expansions of the fluctuations}

In the UV, a general spin-0 fluctuation $\mathfrak{a}(m,z)$ 
can be written as a linear combination of $16$  independent solutions. 
We make a specific choice for such solutions that defines a basis for the vector space of all possible
solutions to the homogeneous second-order linear equations obeyed by the fluctuations.
We split them into two groups
denoted $\mathfrak{a}^{(UV)a}_i$ and $\tilde{\mathfrak{a}}^{(UV)a}_i$, with $i = 1\,,\, \cdots\,,\, 8$
(not to be confused with the sigma-model index  $a=1\,,\,\cdots\,,\,8$),
and we report their detailed structure in Appendix~\ref{Sec:UV}.
The general fluctuation is of the form:
\beq
	\mathfrak a^a(\rho)=  c_i \mathfrak a^{(UV)a}_i(\rho) + \tilde c_i \mathfrak {\tilde a}^{(UV)a}_i(\rho)\,.
	\label{Eq:fluctuation}
\eeq
Out of (the dominant) $\mathfrak {\tilde a}^{(UV)a}_i$, there is one each that starts at orders $z^{-4}$, $z^{-3}$, and $z^{-2}$, 
two each that start at orders $z^0$ and $z^1$, and finally one that starts at order $z^2$ 
(we ignore the $\log(z)$ terms in this rough classification). Out of the subdominant 
$\mathfrak a^{(UV)a}_i$, one starts at order $z^2$, two each at orders $z^3$ and $z^4$, and 
finally one each at orders $z^6$, $z^7$, and $z^8$. These powers reflect the dimensionality of the 
operators in the dual field theory~\cite{dimension}.

Imposing the eight boundary conditions in  Eq.~\eqref{Eq:BCb} at $\rho = \rho_U$, 
the coefficients $c_i$ and $\tilde c_i$ become dependent on $\rho_U$. 
In the limit of $\rho_U \rightarrow \infty$,  we write the coefficients as a power expansion in the form
\beqs
	c_i &=& c^{0}_i + c^{1}_i z_U + \ldots , \\
	\tilde c_i &=& \tilde c^{0}_i + \tilde c^{1}_i z_U + \ldots ,
\eeqs
 where $z_U = e^{-2\rho_U/3}$.
We then expand the boundary conditions in powers of $z_U$, to obtain eight constraint equations for 
$c^{0}_i$ and $\tilde c^{0}_i$. Our choice of linearly independent solutions in Eq.~(\ref{Eq:fluctuation})
is adapted to yield
\beq
	\lim_{z_U \rightarrow 0} \tilde c_i^a(z_U) = 0\,.
\eeq
The conclusion of this analysis is that the $\r_U\rightarrow +\infty$ limit of our boundary conditions
is equivalent to imposing that in the UV
\beq
	\mathfrak a^a(\rho)=  c_i \mathfrak a^{(UV)a}_i(\rho),
	\label{Eq:UV}
\eeq
with arbitrary (real) constants $c_i$. This shows explicitly tha  our procedure is equivalent to the 
conventional wisdom about gauge/gravity dualities, effectively suppressing the dominant fluctuations
(interpreted in terms of couplings in the dual field theory) in respect to the subdominant ones (interpreted as fluctuations of the vacuum value of the field-theory operators).

In the IR, we perform the same exercise. We write a general spin-0 fluctuation as
\beqs
	\mathfrak a^a(\rho)=  d_i \mathfrak a^{(IR)a}_i(\rho) + \tilde d_i \mathfrak {\tilde a}^{(IR)a}_i(\rho)\,,
\eeqs
and we report the explicit form of  $\mathfrak a^{(IR)a}_i$ and $\tilde{\mathfrak a}^{(IR)a}_i$
in Appendix~\ref{Sec:IR}.

Out of $\mathfrak {\tilde a}^{(IR)}_i$, there is one each that starts at orders $\rho^{-3}$ and $\rho^{-2}$, 
there are five that start at order $\rho^{-1}$, and one that starts at order $\rho^0$. 
Out of $\mathfrak a^{(IR)}_i$, there are three that start at order $\rho^0$, one that starts at order $\rho$, and four that start at order $\rho^2$.

Similar to the case of the UV, we expand the coefficients as
\beqs
	d_i &=& d^0_i + d^1_i \rho_I + \ldots , \\
	\tilde d_i &=& \tilde d^0_i + \tilde d^1_i \rho_I + \ldots\,,
\eeqs
we impose the boundary conditions in Eq.~\eqref{Eq:BCb}, and we take the limit $\rho_I \rightarrow 0$.
This process leads to $8$ constraint equations for the coefficients, implying that $\tilde d_i = 0$,
and that
\beq
	\mathfrak a^a(\rho)=  d_i \mathfrak a^{(IR)a}_i(\rho),
	\label{Eq:IR}
\eeq
with arbitrary constants $d_i$.
The analysis of the divergences alone would have left an ambiguity in the choice of 
what fluctuations to suppress. This exercise is equivalent to the standard process of imposing regularity (when 
this is a well defined concept)
in the IR on the fluctuations, with the practical advantage that 
the boundary conditions automatically select the least divergent fluctuations, as they contain the information about the
metric and the kinetic terms in the action.

\section{Mass spectrum}
\label{Sec:spectrum}

As is clear from the asymptotic expansions, the numerical study we 
perform involves large exponential hierarchies between dominant and subdominant contributions
to the solutions of the differential equations, both for the background and for the fluctuations,
in the presence of non-trivial mixing between the eight scalars.
This presents a challenge for the numerical implementation of the procedure to compute the spectrum,
 in particular it limits our practical ability to reach high enough UV cutoff $\r_U$.
 We start this section by explaining in detail the systematic process we employ to address this
  technical problem, before presenting and discussing the results.

We construct the background solutions by setting up the boundary conditions for $P$ in the IR, according to Eq.~\eqref{Eq:baryonic}, 
and solving Eq.~\eqref{Eq:master} numerically. For convenience, we introduce the variable $\alpha$,
 defined in terms of the parameter $\kappa_1$ (appearing in the IR expansion of $P$) as
\beqs
	\kappa_1 &\equiv& 2 + e^{-\alpha}\,.
\eeqs
We evolve the solutions up to a scale $\r_m$ in the radial direction,
beyond which we use the UV expansion of the background functions, by fixing the value of the integration constants
such as $c_+$. In order to do so, we must require that $\r_m>\bar{\r}$, as the UV expansion
is valid only when $P\sim e^{\frac{4}{3}\r}$.

The scale $\bar \rho$, above which $P$ starts to grow exponentially, can be roughly estimated by asking at which 
value of $\rho$ the UV expansion of $P$ starts to break down, i.e. it becomes the same order as the
 linear behaviour $P \simeq 2\rho$. Hence, for $c_+ < \frac{1}{2e}$, we identify $\bar \rho$ 
 as the larger of the two solutions to the equation $3c_+ e^{4\rho/3} = 2\rho$. For $c_+ = \frac{1}{2e}$, this equation has a single solution given by $\bar \rho = \frac{3}{4}$, while for $c_+ > \frac{1}{2e}$ there are no solutions. We fix $\bar \rho = \frac{3}{4}$ also in the latter case, ensuring continuity of $\bar \rho$ as a function of $c_+$. We found that defining $\bar \rho$ in this way to be convenient for the numerics (although note that it is different from the estimate mentioned in Section~\ref{Sec:bg}).

We replace the use of Eq.~(\ref{Eq:BCb}) by making use of Eq.~(\ref{Eq:UV}) and Eq.~(\ref{Eq:IR}). 
Hence, we solve the bulk equations for the fluctuations  of the scalars subject to the boundary conditions
obtained from the asymptotic expansions (and their derivatives), for eight linearly independent choices 
of the functions $\mathfrak{a}^a$ controlled by the $c_i$ coefficients and eight determined by the $d_i$ coefficients.
We impose the boundary conditions at finite values of $\r_I$ and $\r_U$,
with the physical results recovered for $\r_I\rightarrow 0$ and $\r_U\rightarrow +\infty$. 
Had we used Eq.~(\ref{Eq:BCb}), one would expect that the results of the numerics be affected by spurious 
unphysical corrections in the IR and UV. The use of the asymptotic expansion reduces the size of these effects significantly, and results in much faster convergence of the spectrum as $\rho_I \rightarrow 0$ and $\rho_U \rightarrow +\infty$. Hence, our results are close to the physical ones, with negligibly small spurious effects, in spite of the fact that we will not be able to set the boundary conditions at very small (large) values of the radial direction in the IR (UV). This process is indeed  very similar to the {\it improvement} procedure that is common place in the lattice literature in order to remove finite-size effects~\cite{improvement}.

We evolve the scalar fluctuations numerically from the IR and UV, respectively, having imposed the boundary conditions 
deduced from Eq.~(\ref{Eq:UV}) and Eq.~(\ref{Eq:IR}), 
and match them by computing the so-called midpoint determinant~\cite{BHM2} at an intermediate value of $\rho = \rho_{mid}<\bar{\r}<\r_m$. More precisely, one forms the $16 \times 16$ matrix
\beq
	\mathcal M(\rho) =
	\left(
\begin{array}{cc}
 \mathfrak a^{(IR)}(\rho) & \mathfrak a^{(UV)}(\rho) \\
 \partial_\rho \mathfrak a^{(IR)}(\rho) & \partial_\rho \mathfrak a^{(UV)}(\rho)
\end{array}
\right) \, ,
\eeq
where $\mathfrak a^{(IR)}$ ($\mathfrak a^{(UV)}$) is an $8 \times 8$ matrix obtained by lining up next to each other the column vectors corresponding to eight linearly independent solutions that satisfy the boundary conditions in the IR (UV). When $\det \mathcal M = 0$, there exists a linear combination of the solutions evolved from the IR that can be written as a linear combination of those evolved from the UV, and hence Eq.~\eqref{Eq:diffeq} can be solved while satisfying both the IR and UV boundary conditions.

For $\r_I<\r<\r_{mid}$ we use of the numerical solutions for the background
in Eq.~(\ref{Eq:ST}).
The fluctuations are evolved from the UV in two steps.  
In the region $\rho_m \leq \rho \leq \rho_U$, we make use of the UV expansion of the background functions in order to expand 
$S^a_{\ b}$, $T^a_{\ b}$, and $e^{-2A-8p}$ appearing in the equations of motion for the scalar
 fluctuations.\footnote{We here use the UV expansion of $P$ to order $z^{14}$.}
For $\rho_{mid} \leq \rho < \rho_m$, the equations of motion are obtained from the numerical solution of $P$, after joining smoothly the fluctuations at $\r_m$.

For the tensorial fluctuations, the behavior of the spectrum is much less complicated and affected by smaller 
boundary effects, hence we use directly the boundary conditions 
at finite cutoff, without making use of the asymptotic expansions of the fluctuations.
 In the numerical study, we use $\rho_I = 10^{-3}$, $\rho_{mid} = 2 \bar \rho / 3$, 
$\rho_m = \bar \rho + 5$, and $\rho_U = \bar \rho + 10$.
 We have performed the necessary tests to confirm that indeed this ensures that the results of the computation have converged.

Before proceeding to discuss the spectra of the baryonic branch solutions, in the next subsections,
we report here some important results for the KS and CVMN backgrounds.
We performed the calculation of the spectra with the full sigma-model with eight scalars,
rather than within the truncations discussed in the literature (for example in~\cite{BHM1,BHM2}).
The very existence of the sub-truncations means that
the additional states we find are not affected by mixing with the rest of the spectrum.

For the CVMN solution, 
we show explicitly the UV expansion of the fluctuations in Appendix~\ref{Sec:CVMN}.
The interesting fact that emerges is that because of the asymptotics, the 2-point correlation functions
for the tensors contain factors of $\sqrt{1 - m^2/X^2}$, where $X$ depends on the integration constants in the background.
As a consequence, the spectrum has a continuum cut opening up at $m=X$. The spectrum of tensors does not have
any bound states, but only a continuum for $m>X$.
The scalar spectrum is more complicated, and admits infinitely many bound states.
The asymptotic expansion shows the appearance of terms depending on
$\sqrt{1-m^2/X^2}$, $\sqrt{4-m^2/X^2}$, and $\sqrt{9-m^2/X^2}$.
Hence, besides the discrete spectrum, there are also three thresholds for three distinct continuum 
parts of the spectrum, at $m=X$, $m=2X$ and $m=3X$.
Notice that the second of these thresholds is not found in the truncation to six scalars of the sigma-model, while only the first threshold remains in the truncation to three scalars.

For the KS solution, the fact that we use the eight scalar truncation 
implies that there is one additional tower of states. 
We verified explicitly that our results agree with~\cite{BHM2}, except for the presence of this additional 
tower of states in the scalar part of the spectrum (see Appendix~\ref{Sec:CVMN}).

\subsection{Tensorial modes}
\label{Sec:Tensors}

\begin{figure}[h]
\begin{center}
\includegraphics[height=7.1cm]{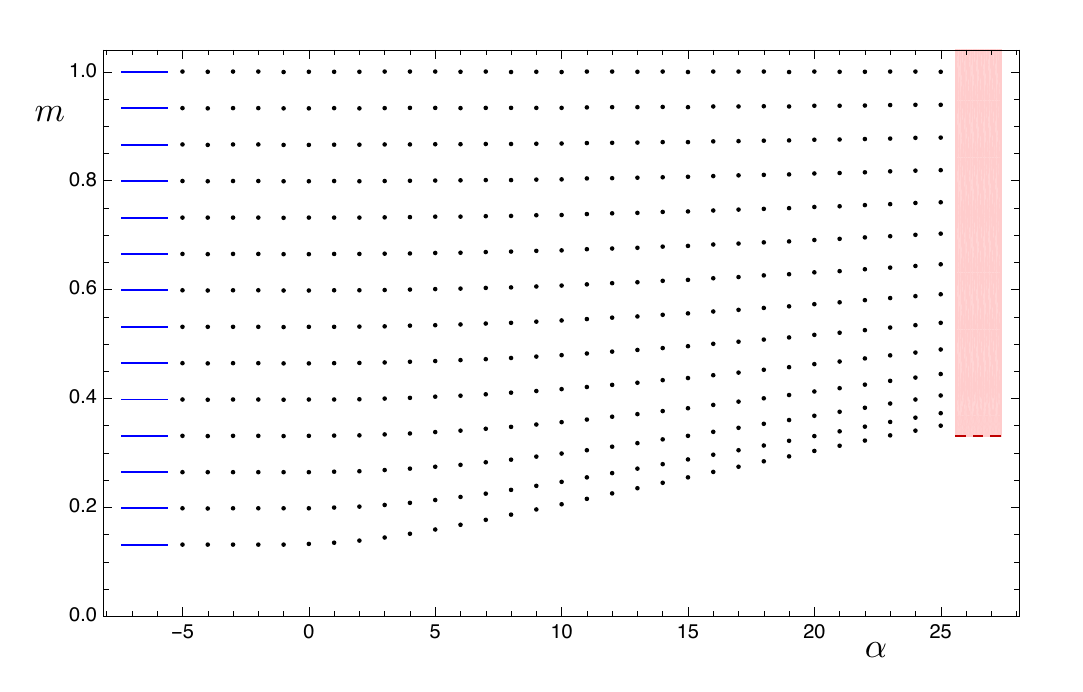}
\caption{Mass spectrum $m$ of tensor modes for different baryonic branch solutions constructed by changing $\alpha$.
The normalization is discussed in the main text. By way of comparison, we show also the KS spectrum (in blue, far left) and the CVMN spectrum (in red, far right). Notice that the latter has a threshold above which a continuum appears (shaded region).}
\label{Fig:tensorsbaryonic}
\end{center}
\end{figure}

\begin{figure}[h]
\begin{center}
\includegraphics[height=16.5cm]{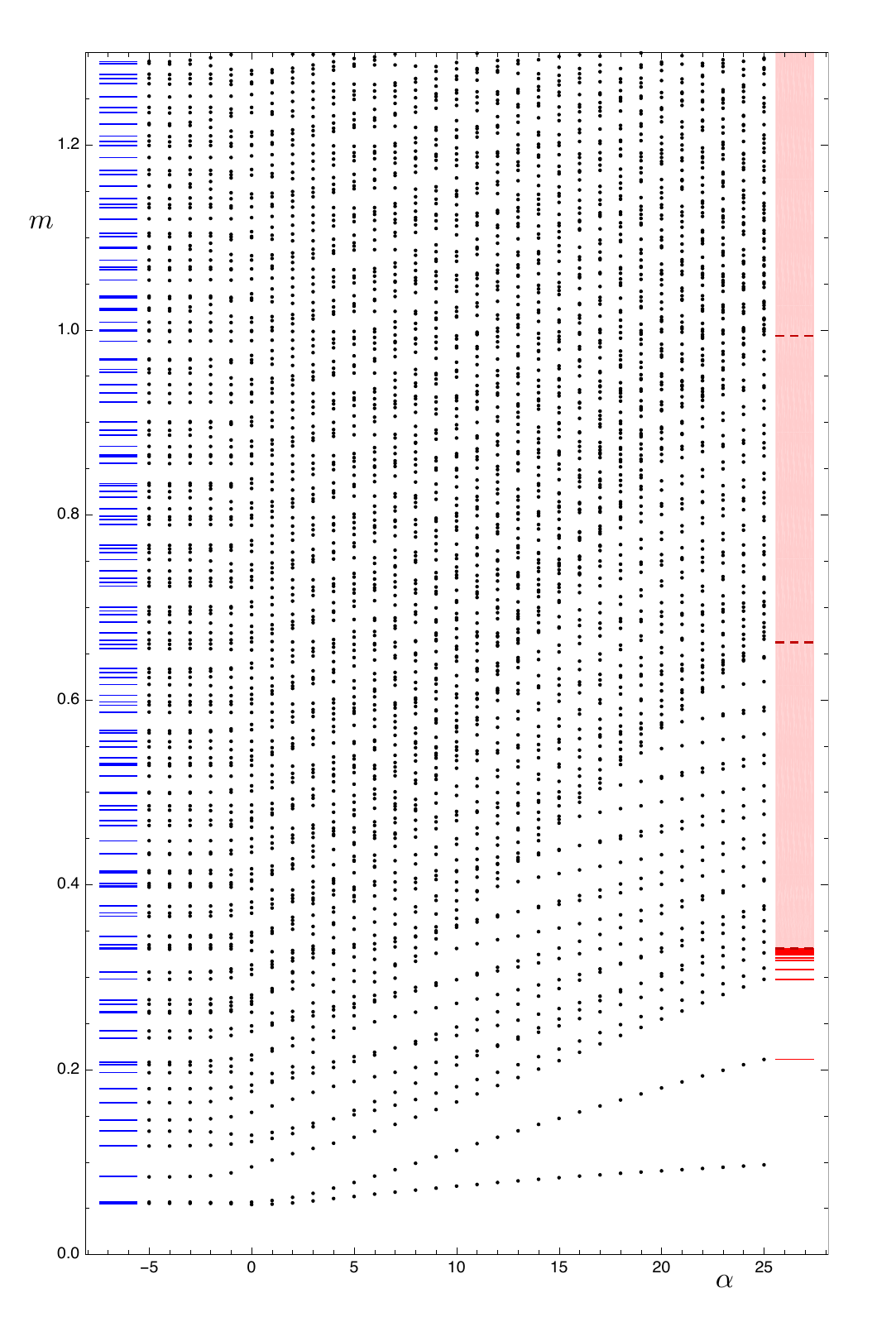}
\caption{The mass spectrum $m$ of scalar modes as a function of the parameter $\alpha$ characterizing the position 
along the baryonic branch of KS. By way of comparison, we show also the KS (in blue, far left) and CVMN limits (in red, far right), defined as in Appendix E. Notice that the latter has continuum thresholds (dashed lines), and that all calculations have been performed within the eight-scalar sigma-model truncation (PT ansatz).}
\label{Fig:scalarbaryonic}
\end{center}
\end{figure}

We start the presentation of the physics results with the tensor modes, the physical spectrum of which is shown in Fig.~\ref{Fig:tensorsbaryonic}.
For each background  (characterized by a different value of $\kappa_1$, or equivalently $\alpha$)
we compute the first fourteen values of the mass $m$.
In order to compare the results for different backgrounds, we normalize the individual spectra
by showing the ratio of the masses to the heaviest mass we found. Another way of saying this is that we always choose the integration constant $A_0$ so that the mass of this state is equal to 1. Note that this implies that the spacing between the heaviest states is the same as in the KS case, thus ensuring that the dynamical scale $\Lambda$ is kept fixed as $\alpha$ is varied.

The comparison shows several non-trivial features. Firstly, the solutions with smallest values of 
$\alpha$ agree with the result of the spectrum of KS computed in the literature~\cite{BHM2}.
We find a rough approximation, in this choice of normalization,  to be
$m_n\simeq \frac{1}{15}(1+n)$, with $n=1\,,\,2\,,\,\cdots$.
For large values of $\alpha$,  only the first few lightest states become heavier, 
while the heavy masses maintain agreement with the KS case.
The resulting distortion of the spectrum shows the appearance of an anomalously dense number of closely-spaced
mass eigenvalues, for $m^2$ over a finite range that grows with $\alpha$.

These results are in qualitative agreement with what is expected on the baryonic branch: while at high energy the 
gravity background and the dual field theory are almost indistinguishable from the KS case,
the lowest states have the closely packed qualitative behavior expected from deconstruction.
In our numerical study, we limited our exploration to $\alpha<25$, but extrapolating the current results
will eventually show a larger and larger number of such densely packed states, until one ultimately
would recover the continuum spectrum of the CVMN case above the threshold.

The field theory interpretation of the baryonic branch perturbative spectrum, as we recalled earlier, shows that 
the number of states is controlled by the parameter $q$, describing  the $SU(q N)\times SU((q+1)N)\rightarrow SU(N)$ coset.
Fig.~\ref{Fig:tensorsbaryonic} shows that $\alpha$ is related to $q$ in a non-perturbative calculation, as by considering solutions with large $\alpha$ yields a spectrum that contains an increasing number of densely packed mass eigenstates.

\subsection{Scalar modes}
\label{Sec:Scalars}

For the scalars, we use the same normalization of the mass spectrum as obtained from the tensor modes,
in such a way that the two plots in Fig.~\ref{Fig:tensorsbaryonic} and Fig.~\ref{Fig:scalarbaryonic} can be compared directly.
Because there are 
eight sigma-model scalars, the spectrum consist of several towers of states, that when superimposed make it 
visually more difficult to recognize the regular patterns emerging.
For small values of $\alpha$, we checked explicitly that our results reproduce those of the KS case (see~\cite{BHM2}
and Appendix~\ref{Sec:CVMN}).

For large values of $\alpha$, the spectrum shows the same type of deformation we observed for the 
tensor modes. At asymptotically large $m^2$, the spectrum maintains its agreement with the KS case,
but the lightest modes are shifted to larger values of $m^2$.
By looking at the largest values of $\alpha$ we were able to study in Fig.~\ref{Fig:scalarbaryonic},
one sees the emergence of three separate values of $m$ (thresholds) at which the discrete spectrum becomes dense.
One also notices that the lightest such threshold agrees numerically with the tensor one,
while the other two are approximately $2$ and $3$ times heavier.
This was expected on the basis of the considerations we made about the CVMN spectrum: three distinct thresholds 
appear in connection with the asymptotic expansion of the two-point functions.

There is an additional element emerging, again expected: contrary to the tensor case, not all the discrete 
states we find are going to merge into the continuum thresholds in the $\alpha\rightarrow +\infty$ limit,
because the scalar spectrum of the CVMN solution does not consist only of continuum cuts, 
but also includes a discrete, infinite set of bound states that appear below the first  threshold.
This is well known and established in the literature on the CVMN spectrum~\cite{BHM1}.
For example, we checked explicitly that, for the largest available values of $\alpha$, the second to fourth lightest states match with the first to third lightest states cited for the CVMN case in~\cite{BHM1}.

The most interesting thing is that there is an additional state in the spectrum in Fig.~\ref{Fig:scalarbaryonic} that cannot be
identified with any of the mass eigenstates of neither the KS nor the CVMN system. This is the lightest mode 
in the spectrum. 
In the small-$\alpha$ case, it is approximately degenerate with one of the KS mass eigenvalues,
but is not listed in~\cite{BHM1}, indicating that it must have its origin in the presence of one additional sigma-model scalar
in the truncation we use for the baryonic branch solutions. Therefore, for $\alpha\rightarrow -\infty$ the eigenstate
it corresponds to is mostly a combination of fluctuations of $\tilde{g}$ and $a$. Moving towards larger values of $\alpha$, this state becomes parametrically lighter than the rest of the spectrum. We expect this state to become exactly massless in the CVMN limit, in the sense that the ratio of its mass to that of any of the other states vanishes for $\alpha \rightarrow +\infty$.

The presence of this parametrically light scalar is the main result of this paper. As we will discuss, it opens great opportunities in the context of field theory, model building and phenomenology, but its existence also opens a few additional questions. We will summarize and discuss these points in the next sections.

We close this subsection with a set of technical remarks. It is tempting to ask what is the general form of the eigenstate describing the light scalar, both in terms of sigma-model and gravity components, and of the $\r$-dependence of the associated wave-function. This is a numerically challenging question. In Eq.~(\ref{Eq:a}), the gauge-invariant combinations of the scalar fluctuations are written in terms of the fluctuations $\varphi^a$ of the individual sigma-model scalars and of the scalar component $h$ of the fluctuations of the metric. The mixing is controlled by the value of the derivative of the sigma-model fields in the background. For generic $\alpha$ and  $\r$, none of the background scalars is constant, and hence all the scalar fluctuations mix with $h$, which in turn means that all the $\mathfrak{a}^a$ mix with one another.

Although one might expect the analysis to become simplified in the strict $\alpha \rightarrow \infty$ limit given by the CVMN background, there are a number of subtleties that arise. On the baryonic branch, the boundary conditions for the fluctuations are set up in the far-UV at $\r_U \gg \bar \r$, and in the evolution towards the IR, non-trivial mixing occurs leading to a very particular set of modes being turned on as one enters the region $\r \lsim \bar \r$ (where the solution is approximated by the CVMN one). In other words, the $\alpha \rightarrow \infty$ limit needs to be taken with caution and with special regard to the form of the boundary conditions for the fluctuations in the UV. Indeed, the fact that for $\r>\bar{\rho}$ the behavior of the background changes is what makes this light state physical (normalizable, in the familiar jargon of gravity dualities), which it would not be otherwise. This is a crucial, technical observation: the reason why the state is retained in the spectrum is related to the fact that the UV asymptotic behavior of all the backgrounds we consider is  dual to the duality cascade.

\section{``Who ordered that?''}
\label{Sec:who}

The emergence, from a technically convoluted calculation such as the one performed 
and presented here, of such a striking result as the presence of one parametrically light 
particle in the otherwise massive and complicated mass spectrum, 
strongly suggests the existence of an elegant and  simple symmetry-based argument 
to explain it.
We outline the argument that lead us to believe that the symmetry reason
behind this result is the spontaneous breaking of scale invariance, and hence to the interpretation
of the light scalar as a \mbox{(pseudo-)dilaton}. 
But we also critically discuss the current limitations of such argument, and 
a set of additional (highly non-trivial) calculations that would be worth pursuing in the 
future in order to put this interpretation on rigorous and firm grounds.

The  clear and well established part of the argument has to do with the internal, abelian global
symmetries of the system. The  spontaneous breaking of $U(1)_B$  leads to the presence of a 
Goldstone boson, as was anticipated in~\cite{GHK}, at a time when the baryonic branch solution was not known explicitly. 
This (pseudo-scalar) state is exactly massless, 
and hence cannot be the light scalar state discussed in this paper, nor its scalar superpartner. 
The authors of~\cite{GHK} also discuss the fact that 
along the baryonic branch the  supersymmetric partner of such Goldstone boson 
should be massless as well, although they hint at a potential
normalizability problem.

The anomalous $U(1)_R$ breaks to $Z_{2N}$, and the gaugino condensate further breaks it spontaneously to
a $Z_2$. The presence of the anomaly has a very important role 
in the whole construction of the field theory and its gravity dual, as it is the distinctive feature that makes 
the KS system so interesting in the context of gauge/gravity dualities. Because the field theory has ${\cal N}=1$ supersymmetry,
the $U(1)_R$ symmetry is bundled together (by the supersymmetry algebra) with dilatation symmetry.
Indeed, in the (anomaly-free) $SU(M)\times SU(M)$ case the  field theory is the Klebanov-Witten~\cite{KW} CFT, 
the gravity dual of which has the geometry of $AdS_5\times T^{1,1}$. 
The presence of the anomaly ensures that the gauge theory is not a CFT, and indeed the two gauge couplings run~\cite{S}.
But in the present case there appears to be a sense in which the anomaly is {\it small}.
If one has a large VEV breaking spontaneously scale invariance,
and a small explicit symmetry breaking effect, one expects a light scalar particle in the spectrum, the dilaton,
and we want to interpret the light scalar that emerged from the calculation performed here
as such a state.

The non-trivial part of this line of thought requires giving a rigorous meaning to the 
qualifier {\it small} as used in the previous paragraph in  the statement that there is a large VEV and a small anomaly.
And in principle one needs to clarify what is the interplay between the non-linear realization of both the $U(1)_B$ and $Z_{2N}$
symmetries, the associated \mbox{(pseudo-)Goldstone} modes, and their supersymmetric partners, 
in order to genuinely understand in what sense the light scalar
particle we found is to be interpreted as a dilaton.

In the presence of a classical moduli space, namely when a given field theory admits a set of non-trivial, inequivalent vacua,
it is not surprising to find evidence of a massless mode
corresponding to excitations of the vacuum along the corresponding flat directions.
When the presence of an anomaly lifts the degeneracy of vacua in the moduli space, this should result in the 
massless mode acquiring a mass. This effect persists in the large-$N$ limit:
the very fact that the study of the gravity dual 
is interpreted in terms of a  supersymmetric field theory in which the couplings 
run means that both scale invariance and the $U(1)_R$ are explicitly broken.

In this paper, when presenting the results for the spectrum, we have been careful to normalise the 
masses so that the heavy states agree. All gravity backgrounds correspond to field theories 
that have the same dynamically generated scale $\Lambda$. If this were the end of the story, there would be no actual sense in which one could give any meaning to statements about the anomaly being small,  or any condensates being large. As happens for example in SYM (and in the KS case), for which we can make use of perturbative 
arguments to guide us: all the physical scales are  controlled by the scale anomaly (beta functions).
The scale of explicit breaking of scale invariance in SYM is determined by the beta function via 
dimensional transmutation, and is the dynamical scale of the theory $\Lambda$. 
The condensates, that introduce spontaneous breaking of scale invariance,
are themselves controlled by the same $\Lambda$.
As a consequence, all the masses of the bound states are controlled by the one and only scale in the problem,
that arises from the beta function via dimensional transmutation, and there is no sense in which a parametrically light dilaton exists.

In the theory we are discussing here, there is an element of novelty.
As in SYM, we can still associate the confinement scale $\Lambda$ with the scale of explicit symmetry breaking
originating from the non-trivial beta functions of the theory (barring the subtleties related to the cascade of Seiberg dualities).
And we still have a limiting case in which the only non-trivial condensate is the gaugino condensate ${\cal O}(\Lambda^3)$:
the KS solution, in which no light scalar state appears in the spectrum.
But when we move away from the origin of the baryonic branch, we are introducing an additional, non-trivial condensate,
the VEV ${\cal U}$, that introduces a new, non-trivial scale controlled by $q$, determining the dimension of the coset in the gauge theory~\cite{MM}, and hence the mass of the heaviest states that decouple
because of the Higgs phenomenon.

In this sense, ${\cal U}$ plays the role of a tunable VEV, that breaks spontaneously scale invariance
at a scale that can be made parametrically large in comparison to the scale of explicit breaking $\Lambda$.
The mass gap of the theory is controlled by $\Lambda$, and hence all the masses of the particles in the system 
are going to be proportional to $\Lambda$. Yet, the pseudo-Goldstone boson associated with dilatations (the dilaton)
must have a mass further suppressed by some power of the ratio between 
explicit and spontaneous breaking. We hence expect $m_0^2\propto \left(\frac{1}{q}\right)^{\gamma} \Lambda^2$.
The fact that there are two scales is also evident from the emergence of deconstruction: 
the spectrum of heavy states does not consist of just bunches of equally spaced massive modes, but
rather the first few states show that their separation is parametrically small compared to the overall scale $\Lambda$.

There are a set of questions that this argument, based upon our current knowledge of the dynamics of the system,
cannot easily answer.
First of all, the emergence of the deconstructed spectrum at large values of $\bar{\r}$, and the fact that
the number of such states grows with $\bar{\r}$, clearly indicate that $q$ and $\bar{\r}$ are monotonically-increasing functions of one another.
But we do not know what the precise relation is.  

Secondly, there appears to be a further complication, due to the fact that 
the VEV of $\cal U$ breaks spontaneously four different symmetries: the baryon $U(1)_B$,
the gauge symmetry $SU(qN)\times SU((q+1)N)$, as well as scale invariance and the $Z_2$.
Furthermore, the theory is supersymmetric, and hence one expects the spectrum to organize itself in supersymmetric multiplets.
As a result, there must exist  non-trivial relations between the Goldstone boson associated with the $U(1)_B$ symmetry,
the dilaton, their superpartners, and the towers of pseudo-scalar and scalar states that, together with the massive (higgsed)
gauge bosons, form massive ${\cal N}=1$ supermultiplets.
For example, it would be nice to know whether a pseudo-scalar partner of the
scalar identified in this paper exists as well, and whether its mass is also
suppressed moving far from the origin of the baryonic branch, a problem that we leave for the future.
The study of the light fermionic states might also be worth pursuing, extending the line of enquiry in~\cite{Argurio}.

In this paper, we studied a special truncation of the gravity dual in which all pseudo-scalar and vector states have been 
removed, and proceeded by brute force to study the spectrum of scalar and tensor fluctations in the resulting sigma model. 
The results we obtained are certainly aligned with the discussion presented in this section. But in order to elucidate, by the same brute force process, the precise relation between the breaking of scale invariance and the internal (global and gauged) symmetries
of the system, one would need  to extend this study to a more general truncation of type-IIB supergravity that includes
the pseudo-scalar and vectorial fields in five dimensions, such as the one in~\cite{CF,BGGHO}, hence also resolving the technical problem with the non-linear constraint discussed in Section~\ref{Sec:5}.
Doing so is quite non-trivial: not only is the scalar manifold in~\cite{CF,BGGHO} significantly more complicated, but the
presence of vectors requires to generalize our whole procedure for treating the fluctuations,
to include in the formalism the effect of gauge invariance not only in the sense of gravity, but also of the internal 
gauge symmetries of such extended 
manifold, as is  clear from the fact that the non-trivial vacuum structure will result in the higgsing of part of the pseudo-scalars
into the massive vector bosons.
We leave this task for a future dedicated study.

\section{Conclusions and Outlook}
\label{Sec:outlook}

In the context of gauge/gravity dualities, we performed the calculation of the mass spectrum of excitations around the backgrounds along the baryonic branch of the KS system, by exploiting the truncation 
to five dimensions corresponding to the PT ansatz, by implementing the gauge-invariant formalism of~\cite{BHM1}
for treating the fluctuations of a sigma-model coupled to gravity, and by imposing the boundary conditions of~\cite{EP}.
We focused our attention on the spin-2 sector of the system and on the spin-0 excitations corresponding to the eight
sigma-model scalars retained in the truncation.

Firstly, we obtained a set of non-trivial, original but expected results. The baryonic branch spectra interpolate between the
known spectra of the KS and CVMN systems, with the additional feature that at finite points along the baryonic 
branch the spectrum shows evidence of the (non-perturbative) emergence of the deconstruction of the internal manifold,
hence confirming earlier  studies based on extrapolating perturbative arguments.
The results show explicitly that the parameter $q$ is related to the gravity scale $\bar{\r}$
that separates the regime over which the backgrounds are approximated by the CVMN solution (for $\r<\bar{\r}$)
and the KS one (for $\r>\bar{\r}$), confirming what was anticipated in~\cite{EGNP}.

We also obtained a remarkable, new and unexpected result.
The scalar spectrum contains one state the mass of which is parametrically suppressed compared to all
others, when moving away from the origin of the baryonic branch.
This state is expected to become massless and completely decouple from the dynamics in the 
limit in which one recovers the CVMN background.

We interpret this state as a dilaton, the pseudo-Goldstone boson associated with 
the spontaneous breaking of scale invariance. We summarized the arguments that lead us to this identification,
as well as  the additional calculations that are required in order to clearly disentangle in the spectrum
the Goldstone boson associated with the spontaneous breaking of $U(1)_B$, the dilaton, and the massive states resulting from the Higgsing 
of the internal gauge symmetries of the system.

This is an example in which the study of the relevant regular backgrounds in the gravity dual supports the existence in the field theory of a non-trivial spectrum exhibiting a parametrically light dilaton, that originates from multi-scale dynamics. The potential implications for phenomenology and model-building in the context of extensions of the standard model are highlighted elsewhere~\cite{elsewhere}. Among the many opportunities opened by this study,
it would be interesting to know whether this model can be used to build semi-realistic, calculable
 models for electroweak symmetry breaking, and whether the results of the model could be generalized to 
 other backgrounds, possibly non-supersymmetric.

\vspace{1.0cm}
\begin{acknowledgments}

We would like to thank D.~Mateos and C.~Nunez for useful discussions, and A.~Faedo for important comments regarding the non-linear constraint of the sigma-model. DE is supported by the ERC Starting Grant HoloLHC-306605 and by the grant MDM-2014-0369 of ICCUB. The work of  MP is supported in part  by the STFC grant ST/L000369/1.

\end{acknowledgments}

\appendix

\section{The KS and CVMN solutions}
\label{Sec:KSCVMN}

We report here the form of all the background functions in the KS and CVMN solutions, 
as useful comparison with 
 the baryonic branch solutions.

The background functions in the KS case are given by the following.
\beqs
	\tilde{g}(\rho) &=& \log (\tanh (2 \rho )) , \\
   	p(\rho) &=& \frac{1}{6} \log \left(\frac{3(\sinh (4 \rho )-4 \rho )}{4\sinh^2(2 \rho )} \right)-\frac{x(\rho )}{3} , \\
   	x(\rho) &=& \frac{1}{3} \log (\sinh (4 \rho )-4\rho ) + \\ \nonumber
	&& \frac{1}{2} \log \left( \int_\rho^\infty \frac{ (2 \rho \coth (2 \rho )-1) \sqrt[3]{\sinh (4 \rho )-4 \rho }}{8\sinh^2(2 \rho )} \right) , \\
   	\phi(\rho) &=& 0 , \\
	 a(\rho) &=& \frac{1}{\cosh(2\rho)}, \\
	 b(\rho) &=& \frac{2\rho}{\sinh(2\rho)} , \\
	 h_1(\rho) &=& \frac{1}{4} \coth (2 \rho ) (2 \rho  \coth (2 \rho )-1) , \\
	 h_2(\rho) &=& \frac{1-2 \rho  \coth (2 \rho )}{ 4\sinh(2 \rho )} , \\
	 A(\rho) &=& \frac{1}{3} \log \left(\sinh (2 \rho )\right) + \frac{x(\rho)}{3} .
\eeqs

As a side remark, we notice that, within the  subtruncation that yields the KS solution,
a whole one-parameter family of (mildly singular\footnote{In the sense that $R$, $R_{\mu\nu}^2$ both are finite but $R_{\mu\nu\sigma\tau}^2$ diverges in the IR.}) solutions, besides those discussed in this paper, 
can be obtained from the general expression in Eq.~(\ref{Eq:P}),
by  holding fixed  the combination $\tilde c_- = \frac{c_-}{c_+^3}$, and by taking the limit $c_+ \rightarrow \infty$.
By doing so one obtains the dual of the KS field theory deformed by a dim-6 VEV controlled by the parameter $f_0$
in~\cite{EGNP,E2}, with $f_0 = -\frac{\tilde c_-}{384}$. 
$f_0 = 0$ corresponds to the KS background. 

The CVMN background is obtained by putting $\kappa_2 = 0$ and $P(\rho) = 2\rho$. 
In this case, the solution has a linear dilaton, and hence one cannot choose $\phi_o$
according to what we did for all other solutions, but rather we must keep it explicit.
This leads to
\beqs
	\tilde{g}(\rho) &=& \frac{1}{2} \log \left(4 \rho ^2-(1-2 \rho  \coth (2 \rho ))^2\right) , \\
	p(\rho) &=& -\frac{\phi _o}{6} + \frac{1}{24} \log \left(-\frac{32768}{8 \rho ^2-4 \rho  \sinh (4 \rho
   )+\cosh (4 \rho )-1}\right) , \\
   	x(\rho) &=& \frac{\phi_o}{2} + \frac{1}{8} \log \left(\frac{\sinh ^2(2 \rho ) \left(4 \rho ^2-(1-2 \rho 
   \coth (2 \rho ))^2\right)^3}{16384}\right) , \\
   	\phi(\rho) &=& \phi _o-\frac{1}{4} \log \left(\frac{1}{4} \left(4 \rho ^2-(1-2 \rho 
   \coth (2 \rho ))^2\right) \text{csch}^2(2 \rho )\right) , \\
   	a(\rho) &=& b(\rho) = \frac{2\rho}{\sinh(2\rho)} , \\
	h_1(\rho) &=& h_2(\rho) = 0 , \\
	A(\rho) &=& \frac{1}{6} \log \left(\frac{1}{32} \left(-8 \rho ^2+4 \rho  \sinh (4
   \rho )-\cosh (4 \rho )+1\right)\right) .
\eeqs

\section{Asymptotic expansions of the backgrounds}
\label{Sec:expansions}

The general baryonic-branch solution in the UV can be written as an expansion for small $z$ in the following manner.
\beqs
	\tilde{g} &=& -\frac{z^2 (3 \log (z)+1)}{3 c_+}+\frac{z^6 \left(-648 c_+^3+9 \log (z)
   (24 \log (z) (\log (z)+1)+17)+35\right)}{324 c_+^3}
   \ \ \ \ \ \\ && \nonumber
   - \frac{z^8
   \left(3 \log (z) \left(-2304 c_+^3 \log (z)-768
   c_+^3+c_-\right)-1152 c_+^3+c_-\right)}{1728
   c_+^4}+ \mathcal O(z^{10})
   , \\
   	p &=& \left(\frac{1}{24} \log \left(\frac{1}{(12 \log (z)+1)^4}\right)+\log
   (2)\right)+
   \\ && \nonumber
   \frac{z^4 (72 \log (z) (12 \log (z) (16 \log
   (z)+19)+145)+1223)}{6912 c_+^2 (12 \log (z)+1)}+
   \\ && \nonumber
   \frac{z^6 \left(60
   \log (z) \left(3840 c_+^3 \log (z)+832 c_+^3-5 c_-\right)+7808
   c_+^3-45 c_-\right)}{48000 c_+^3 (12 \log (z)+1)} + \mathcal O(z^{8})
   , \\
	x &=& \frac{1}{8} \left(\log \left((12 \log (z)+1)^4\right)+\log (81)-28 \log
   (2)\right)+
   \\ && \nonumber
   \frac{z^4 (72 \log (z) (12 \log (z) (8 \log
   (z)+15)+79)-71)}{2304 c_+^2 (12 \log (z)+1)}+
   \\ && \nonumber
   \frac{z^6 \left(60 \log
   (z) \left(23040 c_+^3 \log (z)+5952 c_+^3-5 c_-\right)+12288 c_+^3+5
   c_-\right)}{24000 c_+^3 (12 \log (z)+1)} + \mathcal O(z^{8})
   , \\
	\phi &=& \frac{z^4 (12 \log (z)+1)}{48 c_+^2}-\frac{z^8 (72 \log (z) (12 \log
   (z) (8 \log (z)+5)+65)+823)}{55296 c_+^4}+
   \\ && \nonumber   
   \frac{z^{10} \left(30 \log
   (z) \left(5760 c_+^3 \log (z)+6528 c_+^3+5 c_-\right)+9216 c_+^3+35
   c_-\right)}{432000 c_+^5}+
   \\ && \nonumber
   \frac{z^{12} \left(12 \log (z) \left(144
   \log (z) \left(48 \log ^2(z)+2 \log
   (z)+57\right)-143\right)-5869\right)}{331776
   c_+^6} + \mathcal O(z^{14})
   , \\
	a &=& 2 z^3-\frac{z^5 (6 \log (z)+2)}{3 c_+}+\frac{2 z^7 (3 \log (z)+1)^2}{9
   c_+^2}+
   \\ && \nonumber   
   \frac{1}{6} z^9 \left(\frac{3 \log
   (z)}{c_+^3}+\frac{1}{c_+^3}-12\right)+ \mathcal O(z^{11})
   , \\
	b &=& -6 z^3 \log (z)-6 z^9 \log (z)-6 z^{15} \log (z)-6 z^{21} \log (z)-6
   z^{27} \log (z) + \mathcal O(z^{33})
   , \\
	h_1 &=& \frac{1}{4} (-3 \log (z)-1)-\frac{z^4 ((3 \log (z)+1) (12 \log
   (z)+1))}{96 c_+^2}+ z^6 \left(-3 \log(z)-\frac{1}{2}\right)+
   \ \ \ \ \ \\ && \nonumber   
   \frac{z^8 (3 \log (z)+1) (216 \log (z) (4
   \log (z) (8 \log (z)+1)+19)+799)}{110592 c_+^4} + \mathcal O(z^{10})
   , \\
	h_2 &=& \frac{1}{2} z^3 (3 \log (z)+1)+\frac{z^7 (3 \log (z)+1) (12 \log
   (z)+1)}{48 c_+^2}+\frac{1}{2} z^9 (9 \log (z)+1)-
   \\ && \nonumber   
   \frac{z^{11} ((3
   \log (z)+1) (216 \log (z) (4 \log (z) (8 \log
   (z)+1)+19)+799))}{55296 c_+^4} + \mathcal O(z^{13})
   , \\
	A &=& - \log (z) + \frac{1}{6} \log \left(-\frac{3}{512} (12 \log (z)+1)\right) +
   \\ && \nonumber   
   \frac{z^4 (72 \log (z) (12 \log (z) (8 \log
   (z)+11)+71)-95)}{6912 c_+^2 (12 \log (z)+1)}+
   \\ && \nonumber   
   \frac{z^6 \left(60 \log
   (z) \left(23040 c_+^3 \log (z)+1152 c_+^3-5 c_-\right)-11712 c_+^3+5
   c_-\right)}{72000 c_+^3 (12 \log (z)+1)} + \mathcal O(z^8)
\eeqs

We write explicitly the IR expansion (small $\r$) of the baryonic branch solutions, as well. We find it convenient to explicitly keep the dependence on $\hat{h}(0)$, the value of $\hat{h}$ at the end of space
in the IR, and on $\kappa_1$ as defined in Eq.~(\ref{Eq:baryonic}). 
\beqs
	\tilde{g} &=& \log (2 \rho )-\frac{4 \left(3 \left(\kappa_1-2\right) \kappa_1+2\right) \rho ^2}{9
   \kappa_1^2}
      \\ && \nonumber   
      +\frac{8 \left(3 \kappa_1 \left(\kappa_1 \left(21 \left(\kappa_1-4\right)
   \kappa_1+76\right)+48\right)-136\right) \rho ^4}{405 \kappa_1^4} 
   \\ && \nonumber   
 -  \frac{128
   \left(9 \kappa_1 \left(\kappa_1 \left(3 \kappa_1 \left(\kappa_1 \left(\kappa_1 \left(155
   \kappa_1-903\right)+1540\right)-40\right)-3896\right)-1296\right)+23680\right) \rho ^6}{382725 \kappa_1^6} 
      \\ && \nonumber   
      + \mathcal O(\rho ^8)
   , \\
	p &=& - \frac{\phi_o}{6} + \frac{1}{24} \left(-4 \log \left(\hat{h}(0) \rho \right)-5 \log
   \left(\kappa_1\right)+15 \log (2)\right)
      \\ && \nonumber   
      -\frac{4 \left(6
   \hat{h}(0) \kappa_1^2+\hat{h}(0)-20\right) \rho ^2}{135 \left(\kappa_1^2
   \hat{h}(0)\right)}+
   \\ && \nonumber
   \frac{8 \left(\hat{h}(0) \left(153 \hat{h}(0)
   \kappa_1^4+6 \left(\hat{h}(0)-350\right) \kappa_1^2-92
   \hat{h}(0)+560\right)+5600\right) \rho ^4}{42525 \kappa_1^4
   \hat{h}(0)^2} + \mathcal O(\rho ^6)
   , \\
   	x &=& \frac{\phi_o}{2} + \frac{1}{8} \log \left(\frac{\kappa_1^5
   \hat{h}(0)^4}{2048}\right)+\log (\rho
   )+\frac{4 \left(\left(3 \kappa_1^2+8\right) \hat{h}(0)-20\right)
   \rho ^2}{45 \kappa_1^2 \hat{h}(0)}-
   \\ && \nonumber
   \frac{\left(8 \hat{h}(0) \left(9
   \hat{h}(0) \kappa_1^4+6 \left(173 \hat{h}(0)-350\right) \kappa_1^2-2896
   \hat{h}(0)+560\right)+44800\right) \rho ^4}{14175 \left(\kappa_1^4
   \hat{h}(0)^2\right)} + \mathcal O(\rho ^6)
   , \\
	\phi &=& \phi_o + \frac{1}{4} \log \left(\frac{32}{\kappa_1^3}\right)+\frac{16 \rho ^2}{9 \kappa_1^2}-\frac{32 \left(15 \kappa_1^2-44\right)
   \rho ^4}{405 \kappa_1^4}
      \\ && \nonumber   
      +\frac{256 \left(999 \kappa_1^4-6732 \kappa_1^2+13120\right)
   \rho ^6}{382725 \kappa_1^6} -
   \\ && \nonumber
   \frac{1024 \left(5832 \kappa_1^6-66573 \kappa_1^4+284720
   \kappa_1^2-416240\right) \rho ^8}{17222625 \kappa_1^8} + \mathcal O(\rho ^{10})
   , \\
	a &=& 1+\left(\frac{8}{3 \kappa_1}-2\right) \rho ^2+\frac{2 \left(\kappa_1 \left(\kappa_1
   \left(75 \kappa_1-232\right)+160\right)+64\right) \rho ^4}{45
   \kappa_1^3}+
   \\ && \nonumber
   \frac{4 \left(\kappa_1 \left(\kappa_1 \left(8640-7 \kappa_1 \left(\kappa_1
   \left(915 \kappa_1-4556\right)+6880\right)\right)+17920\right)+4608\right)
   \rho ^6}{4725 \kappa_1^5} + \mathcal O(\rho ^8)
   , \\
	b &=& 1-\frac{2 \rho ^2}{3}+\frac{14 \rho ^4}{45}-\frac{124 \rho
   ^6}{945}+\frac{254 \rho ^8}{4725}   \\ && \nonumber   
   -\frac{292 \rho
   ^{10}}{13365}+\frac{5657908 \rho ^{12}}{638512875}-\frac{65528 \rho
   ^{14}}{18243225} + \mathcal O(\rho ^{16})
   , \\
	h_1 &=& \frac{2 \sqrt{2}  \rho  e^{2 \phi _o}}{3 \kappa_1^{3/2}}+\frac{32
   \sqrt{2} \left(3 \kappa_1^2+10\right)  \rho ^3 e^{2 \phi _o}}{135
   \kappa_1^{7/2}}-\frac{128 \rho ^5 \left(\sqrt{2} \left(9 \kappa_1^4-21
   \kappa_1^2-196\right)  e^{2 \phi _o}\right)}{2835
   \kappa_1^{11/2}}
   \\ && \nonumber
+   \frac{1024 \sqrt{2} \left(81 \kappa_1^6-837 \kappa_1^4-816
   \kappa_1^2+12400\right)  \rho ^7 e^{2 \phi _o}}{382725
   \kappa_1^{15/2}} + \mathcal O(\rho ^9)
   , \\
	h_2 &=& -\frac{2 \rho  \left(\sqrt{2}  e^{2 \phi _o}\right)}{3
   \kappa_1^{3/2}}+\frac{4 \sqrt{2} \left(21 \kappa_1^2-80\right)  \rho ^3
   e^{2 \phi _o}}{135 \kappa_1^{7/2}} 
      \\ && \nonumber   
      - \frac{4 \rho ^5 \left(\sqrt{2} \left(279
   \kappa_1^4-2688 \kappa_1^2+6272\right)  e^{2 \phi _o}\right)}{2835
   \kappa_1^{11/2}}
   \\ && \nonumber
   + \frac{8 \sqrt{2} \left(10287 \kappa_1^6-180144 \kappa_1^4+951168
   \kappa_1^2-1587200\right)  \rho ^7 e^{2 \phi _o}}{382725
   \kappa_1^{15/2}} + \mathcal O(\rho ^9)
   , \\
	A &=& \frac{2\log(\rho)}{3} + \frac{1}{6} \log \left(\frac{1}{4} \kappa_1^2 \hat{h}(0)\right) -\frac{\left(-42 \hat{h}(0) \kappa_1^2+8 \hat{h}(0)+80\right)
   \rho ^2}{135 \left(\kappa_1^2 \hat{h}(0)\right)}
   \\ && \nonumber
 -  \frac{\left(4 \hat{h}(0)
   \left(333 \hat{h}(0) \kappa_1^4-24 \left(\hat{h}(0)+175\right) \kappa_1^2+16
   \left(23 \hat{h}(0)+70\right)\right)+44800\right) \rho ^4}{42525
   \left(\kappa_1^4 \hat{h}(0)^2\right)} + \mathcal O(\rho ^6)
\eeqs

\section{UV expansions of the fluctuations}
\label{Sec:UV}

As explained in the main body of the text, we write here a special basis for the scalar fluctuations that solve the bulk equations 
for the baryonic branch solutions in the far UV
of the five-dimensional geometry. We order the fluctuations on the basis of their degree of 
divergence, having imposed the boundary conditions in Eq.~(\ref{Eq:BCb}).
For the numerical analysis we have computed the expansions of the fluctuations  retaining all terms up to order $z^8$, 
but for presentation purposes we list here only the first few terms.

We start with the dominant fluctuations, that are suppressed by the UV boundary conditions, and are listed in the following.
All expressions are written in the same basis $\Phi^a$ introduced in Section~\ref{Sec:old}.
\beqs
	\mathfrak {\tilde a}^{(UV)}_1 &=&
z^{-4}
\left(
\begin{array}{c}
 0 \\
 \frac{1}{\frac{1}{12}-\log (z)} \\
 \frac{36}{12 \log (z)-1} \\
 0 \\
 0 \\
 0 \\
 \frac{9}{1-12 \log (z)} \\
 0 \\
\end{array}
\right)
+ z^{-2}
\left(
\begin{array}{c}
 \frac{4 (6 \log (z)+5)}{c_+-12 c_+ \log (z)} \\
 \frac{9}{128} 3^{2/3} m^2 \\
 -\frac{9}{64} 3^{2/3} m^2 \\
 0 \\
 0 \\
 0 \\
 \frac{27}{256} 3^{2/3} m^2 \\
 0 \\
\end{array}
\right)
+ z^{-1}
\left(
\begin{array}{c}
 0 \\
 0 \\
 0 \\
 0 \\
 \frac{72}{12 \log (z)-1} \\
 \frac{90}{1-12 \log (z)}-18 \\
 0 \\
 \frac{18 (3 \log (z)+2)}{12 \log (z)-1} \\
\end{array}
\right)
+ \mathcal O(z^0)
\nonumber
, \\
	\mathfrak {\tilde a}^{(UV)}_2 &=&
z^{-3}
\left(
\begin{array}{c}
 0 \\
 0 \\
 0 \\
 0 \\
 0 \\
 1 \\
 0 \\
 \frac{1}{4} \\
\end{array}
\right)
+ z^{-1}
\left(
\begin{array}{c}
 0 \\
 0 \\
 0 \\
 0 \\
 -\frac{3}{128} 3^{2/3} m^2 \\
 -\frac{3}{128} 3^{2/3} m^2 (3 \log (z)+1) \\
 0 \\
 -\frac{3\ 3^{2/3} m^2 (6 \log (z)-1)}{1024} \\
\end{array}
\right)
+
\left(
\begin{array}{c}
 0 \\
 0 \\
 0 \\
 4 \\
 0 \\
 0 \\
 -\frac{1}{2} \\
 0 \\
\end{array}
\right)
+ \mathcal O(z)
\nonumber
, \\
	\mathfrak {\tilde a}^{(UV)}_3 &=&
z^{-2}
\left(
\begin{array}{c}
 0 \\
 \frac{1}{\frac{1}{12}-\log (z)}-\frac{3}{2} \\
 \frac{36}{12 \log (z)-1}-3 \\
 0 \\
 0 \\
 0 \\
 \frac{9}{1-12 \log (z)}-\frac{9}{4} \\
 0 \\
\end{array}
\right)
+
\left(
\begin{array}{c}
 -\frac{16-3 \log (z) (108 \log (z)+143)}{c_+-12 c_+ \log (z)} \\
 -\frac{1}{128} 3^{2/3} m^2 (1-12 \log (z)) \\
 -\frac{3}{256} 3^{2/3} m^2 (1-36 \log (z)) \\
 \frac{3}{128} 3^{2/3} m^2 (36 \log (z)+13) \\
 0 \\
 0 \\
 -\frac{3\ 3^{2/3} m^2 (72 \log (z) (2 \log (z)+3)-13)}{2048} \\
 0 \\
\end{array}
\right)
+ \mathcal O(z)
\nonumber
, \\
	\mathfrak {\tilde a}^{(UV)}_4 &=&
\left(
\begin{array}{c}
 0 \\
 1 \\
 -3 \\
 -6 \\
 0 \\
 0 \\
 \frac{3}{2} (3 \log (z)+1) \\
 0 \\
\end{array}
\right)
+ z^2
\left(
\begin{array}{c}
 0 \\
 -\frac{7\ 3^{2/3} m^2 (6 \log (z) (12 \log (z)-7)-1)}{128 (12 \log
   (z)-1)} \\
 \frac{3\ 3^{2/3} m^2 (12 \log (z) (24 \log (z)-35)+5)}{128 (12 \log
   (z)-1)} \\
 -\frac{27}{64} 3^{2/3} m^2 (4 \log (z)+5) \\
 0 \\
 0 \\
 -\frac{3\ 3^{2/3} m^2 (3 \log (z) (48 \log (z) (9 \log
   (z)-8)-115)+22)}{256 (12 \log (z)-1)} \\
 0 \\
\end{array}
\right)
+ \mathcal O(z^3)
\nonumber
, \\
	\mathfrak {\tilde a}^{(UV)}_5 &=&
\left(
\begin{array}{c}
 0 \\
 \frac{1}{\frac{1}{12}-\log (z)} \\
 \frac{36}{12 \log (z)-1} \\
 0 \\
 0 \\
 0 \\
 \frac{9}{1-12 \log (z)}-\frac{9}{2} \\
 0 \\
\end{array}
\right)
+ z^2
\left(
\begin{array}{c}
 \frac{22}{c_+-12 c_+ \log (z)} \\
 -\frac{21\ 3^{2/3} m^2 (13-12 \log (z))}{768 \log (z)-64} \\
 -\frac{9\ 3^{2/3} m^2 (6 \log (z)-11)}{96 \log (z)-8} \\
 \frac{27}{8} 3^{2/3} m^2 \\
 0 \\
 0 \\
 -\frac{9\ 3^{2/3} m^2 (48 (14-9 \log (z)) \log (z)+31)}{256 (12 \log
   (z)-1)} \\
 0 \\
\end{array}
\right)
+ \mathcal O(z^3)
\nonumber
, \\
	\mathfrak {\tilde a}^{(UV)}_6 &=&
z
\left(
\begin{array}{c}
 0 \\
 0 \\
 0 \\
 0 \\
 1 \\
 \frac{9}{4} \\
 0 \\
 -\frac{3}{16} \\
\end{array}
\right)
+z^3
\left(
\begin{array}{c}
 0 \\
 0 \\
 0 \\
 0 \\
 -\frac{(12 \log (z)-1) \left(128-9\ 3^{2/3} c_+ m^2 (4 \log
   (z)-13)\right)}{1536 c_+} \\
 \frac{(12 \log (z)-1) \left(-3^{2/3} c_+ m^2 (3 \log (z) (12 \log
   (z)-83)+70)-160\right)}{1536 c_+} \\
 0 \\
 \frac{480 (4 \log (z)+1)-3^{2/3} c_+ m^2 (9 \log (z) (67-48 (\log (z)-6)
   \log (z))-320)}{6144 c_+} \\
\end{array}
\right)
+ \mathcal O(z^4)
\nonumber
, \\
	\mathfrak {\tilde a}^{(UV)}_7 &=&
z
\left(
\begin{array}{c}
 0 \\
 0 \\
 0 \\
 0 \\
 \frac{1}{12}-\log (z) \\
 \frac{1}{16}-\frac{3 \log (z)}{2} \\
 0 \\
 \frac{7}{64} \\
\end{array}
\right)
+ z^3
\left(
\begin{array}{c}
 0 \\
 0 \\
 0 \\
 0 \\
 \frac{(12 \log (z)-1) \left(512 (3 \log (z)+1)-3^{2/3} c_+ m^2 (12 \log
   (z) (24 \log (z)-61)+815)\right)}{18432 c_+} \\
 \frac{m^2 (12 \log (z)-1) (12 \log (z) (12 \log (z) (12 \log
   (z)-61)+1187)-5183)}{49152 \sqrt[3]{3}} \\
 0 \\
 -\frac{m^2 (24 \log (z) (144 \log (z) (\log (z) (6 \log
   (z)-23)+24)+1663)-20657)}{196608 \sqrt[3]{3}} \\
\end{array}
\right)
\nonumber
\\
&& + \mathcal O(z^4)
\nonumber
, \\
	\mathfrak {\tilde a}^{(UV)}_8 &=&
z^2
\left(
\begin{array}{c}
 \frac{1}{12}-\log (z) \\
 0 \\
 0 \\
 0 \\
 0 \\
 0 \\
 0 \\
 0 \\
\end{array}
\right)
+ z^4
\left(
\begin{array}{c}
 -\frac{1}{512} 3^{2/3} m^2 (144 (\log (z)-2) \log (z)+215) \\
 \frac{(12 \log (z)-1) (16 \log (z)+11)}{576 c_+} \\
 \frac{36 \log (z) (24 \log (z)+37)+589}{1728 c_+} \\
 \frac{1324 \log (z)+507}{576 c_+} \\
 0 \\
 0 \\
 -\frac{448 \log ^2(z)+412 \log (z)+41}{512 c_+} \\
 0 \\
\end{array}
\right)
+ \mathcal O(z^5)
\nonumber .
\eeqs

The fluctuations that are retained as physically relevant, and ultimately contribute to the eigenstates 
associated with the spectrum, are the following.
\beqs
	\mathfrak a^{(UV)}_1 &=&
z^2
\left(
\begin{array}{c}
 1 \\
 0 \\
 0 \\
 0 \\
 0 \\
 0 \\
 0 \\
 0 \\
\end{array}
\right)
+ z^4
\left(
\begin{array}{c}
 -\frac{3}{128} 3^{2/3} m^2 (11-12 \log (z)) \\
 \frac{1-12 \log (z)}{36 c_+} \\
 -\frac{9 \log (z)+14}{18 c_+} \\
 -\frac{132 \log (z)+107}{72 c_+} \\
 0 \\
 0 \\
 \frac{75 \log (z)+14}{48 c_+} \\
 0 \\
\end{array}
\right)
+ z^5
\left(
\begin{array}{c}
 0 \\
 0 \\
 0 \\
 0 \\
 2 \\
 0 \\
 0 \\
 0 \\
\end{array}
\right)
+ \mathcal O(z^6)
\nonumber
, \\
	\mathfrak a^{(UV)}_2 &=&
z^3
\left(
\begin{array}{c}
 0 \\
 0 \\
 0 \\
 0 \\
 0 \\
 1 \\
 0 \\
 -\frac{1}{4} \\
\end{array}
\right)
+ z^5
\left(
\begin{array}{c}
 0 \\
 0 \\
 0 \\
 0 \\
 -\frac{3}{64} 3^{2/3} m^2 \\
 -\frac{3}{512} 3^{2/3} m^2 (1-24 \log (z)) \\
 0 \\
 -\frac{3\ 3^{2/3} m^2 (24 \log (z)+5)}{2048} \\
\end{array}
\right)
+ z^6
\left(
\begin{array}{c}
 0 \\
 \frac{13}{6} \\
 3 \\
 0 \\
 0 \\
 0 \\
 -\frac{1}{4} \\
 0 \\
\end{array}
\right)
+ \mathcal O(z^7)
\nonumber
, \\
	\mathfrak a^{(UV)}_3 &=&
z^3
\left(
\begin{array}{c}
 0 \\
 0 \\
 0 \\
 0 \\
 1 \\
 \frac{1}{4}-3 \log (z) \\
 0 \\
 \frac{3}{16} (4 \log (z)+1) \\
\end{array}
\right)
+ z^5
\left(
\begin{array}{c}
 0 \\
 0 \\
 0 \\
 0 \\
 -\frac{3 \log (z)+1}{3 c_+}-\frac{3}{256} 3^{2/3} m^2 (23-24 \log (z))
   \\
 -\frac{3\ 3^{2/3} m^2 (72 \log (z) (4 \log (z)-1)-191)}{2048} \\
 0 \\
 \frac{9\ 3^{2/3} m^2 \left(96 \log ^2(z)+8 \log (z)-115\right)}{8192} \\
\end{array}
\right)
+ \mathcal O(z^6)
\nonumber
, \\
	\mathfrak a^{(UV)}_4 &=&
z^4
\left(
\begin{array}{c}
 0 \\
 \frac{1}{\frac{1}{12}-\log (z)} \\
 \frac{36}{12 \log (z)-1} \\
 48 \\
 0 \\
 0 \\
 \frac{108 \log (z)}{1-12 \log (z)} \\
 0 \\
\end{array}
\right)
+ z^6
\left(
\begin{array}{c}
 \frac{4 (6 \log (z)+5)}{c_+-12 c_+ \log (z)} \\
 \frac{1}{640} 3^{2/3} m^2 (216 \log (z)+331) \\
 \frac{3}{320} 3^{2/3} m^2 (72 \log (z)+203) \\
 -\frac{3}{8} 3^{2/3} m^2 (1-12 \log (z)) \\
 0 \\
 0 \\
 -\frac{3\ 3^{2/3} m^2 (672 \log (z)+83)}{1280} \\
 0 \\
\end{array}
\right)
+ \mathcal O(z^7)
\nonumber
, \\
	\mathfrak a^{(UV)}_5 &=&
z^4
\left(
\begin{array}{c}
 0 \\
 1 \\
 -1 \\
 -3 (4 \log (z)+3) \\
 0 \\
 0 \\
 \frac{3}{4} (6 \log (z)+1) \\
 0 \\
\end{array}
\right)
+ z^6
\left(
\begin{array}{c}
 \frac{48 \log (z)-5}{6 c_+} \\
 -\frac{m^2 (72 \log (z) (30 \log (z)-89)+5807)}{12800 \sqrt[3]{3}} \\
 -\frac{3\ 3^{2/3} m^2 (16 \log (z) (15 \log (z)-2)+97)}{6400} \\
 -\frac{9}{128} 3^{2/3} m^2 (4 \log (z)-3) (4 \log (z)+5) \\
 0 \\
 0 \\
 -\frac{3^{2/3} m^2 (6 (683-1260 \log (z)) \log (z)+817)}{12800} \\
 0 \\
\end{array}
\right)
+ \mathcal O(z^7)
\nonumber
, \\
	\mathfrak a^{(UV)}_6 &=&
z^6
\left(
\begin{array}{c}
 0 \\
 \frac{1}{\frac{1}{12}-\log (z)}+\frac{45}{2} \\
 \frac{36}{12 \log (z)-1}+45 \\
 0 \\
 0 \\
 0 \\
 \frac{9}{1-12 \log (z)}-\frac{45}{4} \\
 0 \\
\end{array}
\right)
+ z^8
\left(
\begin{array}{c}
 \frac{-540 \log ^2(z)+69 \log (z)+20}{c_+-12 c_+ \log (z)} \\
 -\frac{m^2 (803-1680 \log (z))}{768 \sqrt[3]{3}} \\
 -\frac{m^2 (37-795 \log (z))}{64 \sqrt[3]{3}} \\
 \frac{325}{128} 3^{2/3} m^2 \\
 0 \\
 0 \\
 -\frac{m^2 (6690 \log (z)+281)}{1024 \sqrt[3]{3}} \\
 0 \\
\end{array}
\right)
+ \mathcal O(z^9)
\nonumber
, \\
	\mathfrak a^{(UV)}_7 &=&
z^7
\left(
\begin{array}{c}
 0 \\
 0 \\
 0 \\
 0 \\
 1 \\
 \frac{3}{25} (15 \log (z)-4) \\
 0 \\
 \frac{3}{200} (30 \log (z)+17) \\
\end{array}
\right)
+ \mathcal O(z^9)
\nonumber
, \\
	\mathfrak a^{(UV)}_8 &=&
z^8
\left(
\begin{array}{c}
 0 \\
 -\frac{80 \log (z)}{3}+\frac{1}{\frac{1}{12}-\log (z)}+\frac{134}{9} \\
 80 \log (z)+\frac{36}{12 \log (z)-1}+\frac{106}{3} \\
 100 \\
 0 \\
 0 \\
 -65 \log (z)+\frac{9}{1-12 \log (z)}-\frac{43}{3} \\
 0 \\
\end{array}
\right)
+ \mathcal O(z^9)
\nonumber .
\eeqs

\section{IR expansions of the fluctuations}
\label{Sec:IR}

As for the UV expansion of the fluctuations on the baryonic branch, we report here the special basis 
we employ for the IR expansions. The details on how this basis has been constructed are in the main text.
In the numerical study, we retained all expansions up to  ${\cal O}(\rho^6)$, 
but we report only the first two terms of the expansion in this Appendix.

The fluctuations that are suppressed by the IR boundary conditions, and hence discarded in our analysis,
are the following.

\beqs
	\mathfrak {\tilde a}^{(IR)}_1 &=&
\rho^{-3}
\left(
\begin{array}{c}
 -1 \\
 \frac{1}{2} \\
 1 \\
 0 \\
 0 \\
 0 \\
 0 \\
 0 \\
\end{array}
\right)
+ \rho^{-1}
\left(
\begin{array}{c}
 0 \\
 0 \\
 0 \\
 0 \\
 -\frac{-45\ 2^{2/3} \hat{h}(0)^2 \kappa_1^3 m^2 e^{4 \text{$\phi $0}/3}+96
   \left(\kappa_1+2\right) \left(9 \kappa_1+2\right) \hat{h}(0)+1280}{120 \kappa_1
   \left(3 \kappa_1-4\right) \hat{h}(0)} \\
 0 \\
 0 \\
 0 \\
\end{array}
\right)
+ \mathcal O(\rho^0)
\nonumber
, \\
	\mathfrak {\tilde a}^{(IR)}_2 &=&
\rho^{-2}
\left(
\begin{array}{c}
 0 \\
 0 \\
 0 \\
 0 \\
 0 \\
 0 \\
 -1 \\
 1 \\
\end{array}
\right)
+
\rho^{-1}
\left(
\begin{array}{c}
 0 \\
 0 \\
 0 \\
 0 \\
 -\frac{32}{4 \kappa_1 \hat{h}(0)-3 \kappa_1^2 \hat{h}(0)} \\
 0 \\
 0 \\
 0 \\
\end{array}
\right)
+ \mathcal O(\rho^0)
\nonumber
, \\
	\mathfrak {\tilde a}^{(IR)}_3 &=&
\rho^{-1}
\left(
\begin{array}{c}
 1 \\
 0 \\
 0 \\
 0 \\
 \frac{3 \kappa_1}{3 \kappa_1-4} \\
 0 \\
 0 \\
 0 \\
\end{array}
\right)
+
\left(
\begin{array}{c}
 0 \\
 0 \\
 0 \\
 0 \\
 0 \\
 0 \\
 \frac{10 \sqrt{2} \sqrt{\kappa_1} \hat{h}(0) e^{2 \text{$\phi $0}}}{\left(3
   \kappa_1-4\right) \left(\hat{h}(0)-3\right)} \\
 0 \\
\end{array}
\right)
+ \mathcal O(\rho)
\nonumber
, \\
	\mathfrak {\tilde a}^{(IR)}_4 &=&
\rho^{-1}
\left(
\begin{array}{c}
 0 \\
 1 \\
 0 \\
 0 \\
 \frac{18 \kappa_1}{4-3 \kappa_1} \\
 0 \\
 0 \\
 0 \\
\end{array}
\right)
+
\left(
\begin{array}{c}
 0 \\
 0 \\
 0 \\
 0 \\
 0 \\
 0 \\
 \frac{6 \sqrt{2} \sqrt{\kappa_1} \left(3 \kappa_1-14\right) \hat{h}(0) e^{2
   \text{$\phi $0}}}{\left(3 \kappa_1-4\right) \left(\hat{h}(0)-3\right)} \\
 0 \\
\end{array}
\right)
+ \mathcal O(\rho)
\nonumber
, \\
	\mathfrak {\tilde a}^{(IR)}_5 &=&
\rho^{-1}
\left(
\begin{array}{c}
 0 \\
 0 \\
 1 \\
 0 \\
 \frac{6 \kappa_1}{4-3 \kappa_1} \\
 0 \\
 0 \\
 0 \\
\end{array}
\right)
+
\left(
\begin{array}{c}
 0 \\
 0 \\
 0 \\
 0 \\
 0 \\
 0 \\
 \frac{3 \sqrt{2} \left(\kappa_1-8\right) \sqrt{\kappa_1} \hat{h}(0) e^{2
   \text{$\phi $0}}}{\left(3 \kappa_1-4\right) \left(\hat{h}(0)-3\right)} \\
 0 \\
\end{array}
\right)
+ \mathcal O(\rho)
\nonumber
, \\
	\mathfrak {\tilde a}^{(IR)}_6 &=&
\rho^{-1}
\left(
\begin{array}{c}
 0 \\
 0 \\
 0 \\
 1 \\
 0 \\
 0 \\
 0 \\
 0 \\
\end{array}
\right)
+
\left(
\begin{array}{c}
 0 \\
 0 \\
 0 \\
 0 \\
 0 \\
 0 \\
 \frac{\sqrt{\kappa_1} \hat{h}(0) e^{2 \text{$\phi $0}}}{\sqrt{2}
   \left(3-\hat{h}(0)\right)} \\
 0 \\
\end{array}
\right)
+ \mathcal O(\rho)
\nonumber
, \\
	\mathfrak {\tilde a}^{(IR)}_7 &=&
\rho^{-1}
\left(
\begin{array}{c}
 0 \\
 0 \\
 0 \\
 0 \\
 -\frac{4}{4 \kappa_1 \hat{h}(0)-3 \kappa_1^2 \hat{h}(0)} \\
 1 \\
 0 \\
 0 \\
\end{array}
\right)
+
\left(
\begin{array}{c}
 0 \\
 0 \\
 0 \\
 0 \\
 0 \\
 0 \\
 -\frac{8 \sqrt{2} \left(9 \kappa_1-17\right) e^{2 \text{$\phi $0}}}{3
   \kappa_1^{3/2} \left(3 \kappa_1-4\right) \left(\hat{h}(0)-3\right)} \\
 0 \\
\end{array}
\right)
+ \mathcal O(\rho)
\nonumber
, \\
	\mathfrak {\tilde a}^{(IR)}_8 &=&
\left(
\begin{array}{c}
 0 \\
 0 \\
 0 \\
 0 \\
 0 \\
 0 \\
 \frac{6-\left(3 \kappa_1^2+2\right) \hat{h}(0)}{2 \left(\hat{h}(0)-3\right)}
   \\
 1 \\
\end{array}
\right)
+ \rho
\left(
\begin{array}{c}
 -\frac{\hat{h}(0) \left(9 \kappa_1^2 \left(-5\ 2^{2/3} \kappa_1 \hat{h}(0) m^2
   e^{4 \text{$\phi $0}/3}-68\right)+2528\right)+5680}{30 \kappa_1^2
   \left(\hat{h}(0)-3\right) \hat{h}(0)} \\
 \frac{\left(27 \kappa_1^2-8\right) \hat{h}(0)-340}{15 \kappa_1^2
   \left(\hat{h}(0)-3\right) \hat{h}(0)} \\
 -\frac{9 \hat{h}(0) \left(\kappa_1^2 \left(-5\ 2^{2/3} \kappa_1 \hat{h}(0) m^2
   e^{4 \text{$\phi $0}/3}-24\right)+256\right)+2080}{60 \kappa_1^2
   \left(\hat{h}(0)-3\right) \hat{h}(0)} \\
 -\frac{\hat{h}(0) \left(448-9\ 2^{2/3} \kappa_1^3 \hat{h}(0) m^2 e^{4
   \text{$\phi $0}/3}\right)+768}{6 \kappa_1^2 \left(\hat{h}(0)-3\right)
   \hat{h}(0)} \\
 0 \\
 -\frac{16}{\hat{h}(0)-3} \\
 0 \\
 0 \\
\end{array}
\right)
+ \mathcal O(\rho^2)
\nonumber .
\eeqs

The subleading fluctuations, that effectively implement regularity in the deep-IR of the geometry,
are the following.

\beqs
	\mathfrak a^{(IR)}_1 &=&
\left(
\begin{array}{c}
 1 \\
 0 \\
 \frac{1}{2} \\
 0 \\
 0 \\
 0 \\
 0 \\
 0 \\
\end{array}
\right)
+ \rho^2
\left(
\begin{array}{c}
 0 \\
 0 \\
 0 \\
 0 \\
 \frac{3 \hat{h}(0) \left(-2^{2/3} \hat{h}(0) \kappa_1^3 m^2 e^{4 \text{$\phi
   $0}/3}-512 \kappa_1+256\right)+2560}{96 \left(4-3 \kappa_1\right) \kappa_1
   \hat{h}(0)} \\
 \frac{\hat{h}(0)^2 \kappa_1^3 m^2 e^{4 \text{$\phi $0}/3}}{64
   \sqrt[3]{2}}+\frac{8}{3} \\
 0 \\
 0 \\
\end{array}
\right)
+ \mathcal O(\rho^3)
\nonumber
, \\
	\mathfrak a^{(IR)}_2 &=&
\left(
\begin{array}{c}
 0 \\
 1 \\
 -3 \\
 0 \\
 0 \\
 0 \\
 0 \\
 0 \\
\end{array}
\right)
+ \rho^2
\left(
\begin{array}{c}
 0 \\
 0 \\
 0 \\
 0 \\
 \frac{-7\ 2^{2/3} \hat{h}(0)^2 \kappa_1^3 m^2 e^{4 \text{$\phi $0}/3}-256
   \hat{h}(0)+1024}{64 \kappa_1 \hat{h}(0)-48 \kappa_1^2 \hat{h}(0)} \\
 8-\frac{5 \kappa_1^3 \hat{h}(0)^2 m^2 e^{4 \text{$\phi $0}/3}}{32
   \sqrt[3]{2}} \\
 0 \\
 0 \\
\end{array}
\right)
+ \mathcal O(\rho^3)
\nonumber
, \\
	\mathfrak a^{(IR)}_3 &=&
\left(
\begin{array}{c}
 0 \\
 0 \\
 0 \\
 1 \\
 0 \\
 0 \\
 0 \\
 0 \\
\end{array}
\right)
+ \rho^2
\left(
\begin{array}{c}
 0 \\
 0 \\
 0 \\
 0 \\
 \frac{3\ 2^{2/3} \hat{h}(0)^2 \kappa_1^3 m^2 e^{4 \text{$\phi $0}/3}+256
   \hat{h}(0)+512}{384 \kappa_1 \hat{h}(0)-288 \kappa_1^2 \hat{h}(0)} \\
 \frac{4}{3}-\frac{\kappa_1^3 \hat{h}(0)^2 m^2 e^{4 \text{$\phi $0}/3}}{64
   \sqrt[3]{2}} \\
 0 \\
 0 \\
\end{array}
\right)
+ \mathcal O(\rho^3)
\nonumber
, \\
	\mathfrak a^{(IR)}_4 &=&
\rho
\left(
\begin{array}{c}
 0 \\
 0 \\
 0 \\
 0 \\
 0 \\
 0 \\
 1 \\
 -1 \\
\end{array}
\right)
+ \rho^2
\left(
\begin{array}{c}
 0 \\
 0 \\
 0 \\
 0 \\
 \frac{32}{4 \kappa_1 \hat{h}(0)-3 \kappa_1^2 \hat{h}(0)} \\
 0 \\
 0 \\
 0 \\
\end{array}
\right)
+ \mathcal O(\rho^3)
\nonumber
, \\
	\mathfrak a^{(IR)}_5 &=&
\rho^2
\left(
\begin{array}{c}
 1 \\
 0 \\
 0 \\
 0 \\
 \frac{3 \kappa_1}{3 \kappa_1-4} \\
 0 \\
 0 \\
 0 \\
\end{array}
\right)
+ \rho^3
\left(
\begin{array}{c}
 0 \\
 0 \\
 0 \\
 0 \\
 0 \\
 0 \\
 -\frac{16 \sqrt{2} e^{2 \text{$\phi $0}}}{\left(12-9 \kappa_1\right)
   \kappa_1^{3/2}} \\
 \frac{16 \sqrt{2} e^{2 \text{$\phi $0}}}{\left(12-9 \kappa_1\right)
   \kappa_1^{3/2}} \\
\end{array}
\right)
+ \mathcal O(\rho^4)
\nonumber
, \\
	\mathfrak a^{(IR)}_6 &=&
\rho^2
\left(
\begin{array}{c}
 0 \\
 1 \\
 0 \\
 0 \\
 \frac{12 \kappa_1}{3 \kappa_1-4} \\
 -\frac{3}{2} \kappa_1^2 \hat{h}(0) \\
 0 \\
 0 \\
\end{array}
\right)
+ \rho^3
\left(
\begin{array}{c}
 0 \\
 0 \\
 0 \\
 0 \\
 0 \\
 0 \\
 \frac{99 \kappa_1+\left(3 \kappa_1 \left(\kappa_1 \left(3
   \kappa_1-4\right)-23\right)-68\right) \hat{h}(0)+28}{30 \left(3
   \kappa_1-4\right)} \\
 \frac{8 \left(17 \hat{h}(0)-7\right)+3 \kappa_1 \left(\left(3 \kappa_1 \left(3
   \kappa_1-4\right)+46\right) \hat{h}(0)-66\right)}{60 \left(3 \kappa_1-4\right)}
   \\
\end{array}
\right)
+ \mathcal O(\rho^4)
\nonumber
, \\
	\mathfrak a^{(IR)}_7 &=&
\rho^2
\left(
\begin{array}{c}
 0 \\
 0 \\
 1 \\
 0 \\
 \frac{3 \kappa_1}{4-3 \kappa_1} \\
 \frac{3}{4} \kappa_1^2 \hat{h}(0) \\
 0 \\
 0 \\
\end{array}
\right)
+ \rho^3
\left(
\begin{array}{c}
 0 \\
 0 \\
 0 \\
 0 \\
 0 \\
 0 \\
 \frac{68 \hat{h}(0)+3 \kappa_1 \left(\left(\left(4-3 \kappa_1\right) \kappa_1+3\right)
   \hat{h}(0)-13\right)-28}{60 \left(3 \kappa_1-4\right)} \\
 \frac{-9 \left(\kappa_1 \left(3 \kappa_1-4\right)+2\right) \hat{h}(0) \kappa_1+78
   \kappa_1-136 \hat{h}(0)+56}{120 \left(3 \kappa_1-4\right)} \\
\end{array}
\right)
+ \mathcal O(\rho^4)
\nonumber
, \\
	\mathfrak a^{(IR)}_8 &=&
\rho^2
\left(
\begin{array}{c}
 0 \\
 0 \\
 0 \\
 1 \\
 \frac{3 \kappa_1}{8-6 \kappa_1} \\
 -\frac{3}{8} \kappa_1^2 \hat{h}(0) \\
 0 \\
 0 \\
\end{array}
\right)
+ \rho^3
\left(
\begin{array}{c}
 0 \\
 0 \\
 0 \\
 0 \\
 0 \\
 0 \\
 \frac{4 \left(37 \hat{h}(0)-27\right)+3 \kappa_1 \left(\left(\left(4-3
   \kappa_1\right) \kappa_1-17\right) \hat{h}(0)+7\right)}{120 \left(3
   \kappa_1-4\right)} \\
 \frac{-296 \hat{h}(0)+3 \kappa_1 \left(\left(3 \left(4-3 \kappa_1\right)
   \kappa_1+34\right) \hat{h}(0)-14\right)+216}{240 \left(3 \kappa_1-4\right)} \\
\end{array}
\right)
+ \mathcal O(\rho^4)
\nonumber .
\eeqs

\section{About the spectrum from the CVMN and KS solutions}
\label{Sec:CVMN}

In this Appendix, we report some useful results about the spectrum of spin-0 states in the CVMN and KS
case. The spectrum has been computed before in the case of the 3-scalar sigma-model 
truncation of the CVMN system~\cite{BHM1}.
Here, for comparison with the calculations performed along the baryonic branch, we consider general fluctuations
involving all the eight sigma-model scalars of the PT ansatz.

In the case of the CVMN solution, it turns out that the only thing we  need to explain is related to the UV-expansions of the 
fluctuations, which we report here.
It is convenient to perform a change of variables $\mathfrak a^a = B^a_{\ b} \tilde {\mathfrak a}^b$, where
\beqs
	B =
\left(
\begin{array}{cccccccc}
 0 & -1 & 0 & 0 & 0 & 1 & 0 & 0 \\
 -\frac{1}{6} & 0 & 0 & \frac{1}{2} & 0 & 0 & 0 & 0 \\
 \frac{1}{2} & -\frac{1}{2} & 0 & \frac{1}{2} & 0 & -\frac{1}{2} & 0 & 0
   \\
 1 & 1 & 0 & 1 & 0 & 1 & 0 & 0 \\
 0 & 0 & 1 & 0 & -1 & 0 & 0 & 0 \\
 0 & 0 & 1 & 0 & 1 & 0 & 0 & 0 \\
 0 & 0 & 0 & 0 & 0 & 0 & 1 & 0 \\
 0 & 0 & 0 & 0 & 0 & 0 & 0 & 1 \\
\end{array}
\right) .
\eeqs

\begin{table}[t]
\begin{center}
\begin{tabular}{|c|c|}
\hline\hline
$8$ scalars &  $7$ scalars~\cite{BHM2}\cr
\hline
	$0.185$ & $0.185$  \cr
	$0.195$ &  \cr
	$0.428$ & $0.428$  \cr
	$0.835$ & $0.835$  \cr
	$1.08$ & \cr
	$1.28$ & $1.28$  \cr
	$1.63$ & $1.63$  \cr
	$1.94$ & $1.94$  \cr
	$2.34$ & $2.34$  \cr
	$2.55$ & \cr
	$2.61$ & $2.61$  \cr
	$3.32$ & $3.32$  \cr
	$3.53$ & $3.54$  \cr
	$4.12$ & $4.12$  \cr 
	$4.17$ & $4.18$  \cr
	$4.43$ & $4.43$  \cr
	$4.43$ & $4.43$  \cr
	$4.58$ & \cr
	$5.35$ & $5.35$  \cr
\hline\hline
\end{tabular}
\end{center}
\caption{The scalar spectrum $m^2$ of the KS solution as calculated by us with the eight-scalar sigma-model,
compared to the seven-scalar sigma-model adopted in~\cite{BHM2}. We normalize the values of $m^2$ so that they can be compared. The empty spaces correspond to states that are missing in the case of seven scalars.}
\label{Fig:KS}
\end{table}

In this basis, we obtain to leading order in $e^{-2\rho/3}$ the following UV expansions for the coefficients defined in Eq.~\eqref{eq:SandT} that appear in the equation of motion for the scalar fluctuations:
\beqs
	\tilde S &=& B^{-1} S B =
	\\ && \nonumber
	{\rm diag} \left( \frac{8 \rho}{4 \rho-1},\frac{8 \rho}{4 \rho-1},\frac{4-8\rho}{1-4 \rho},\frac{8 \rho}{4
   \rho-1},\frac{4-8\rho}{1-4 \rho},\frac{8 \rho}{4 \rho-1},\frac{16 (1-2 \rho) \rho}{(4 \rho-1)
   \left(8 \rho^2-4 \rho+1\right)},0 \right) ,
   \\
	\tilde T &=& B^{-1} T B =
	\\ && \nonumber
	{\rm diag} \left(0,\frac{8}{4 \rho-1}-\frac{2}{\rho^2},\frac{4}{4 \rho-1},-\frac{8}{(1-4
   \rho)^2}-8,\frac{4}{1-4 \rho}-8,-8,-\frac{4}{(1-4 \rho)^2}-4,-4\right) ,
\eeqs
while
\beq
	4 e^{-2A-8p} = \frac{e^{4 \phi_o /3}}{2^{4/3}} \equiv \frac{1}{X^2}.
\eeq
From these expressions it can be seen that to leading order the fluctuations behave as
\beq
	\tilde {\mathfrak a}^{1,2,3} \sim e^{\left(-1 \pm \sqrt{1 - m^2/X^2}\right) \rho}, \hspace{0.3cm}
	\tilde {\mathfrak a}^{4,5,6} \sim e^{\left(-1 \pm \sqrt{9 - m^2/X^2}\right) \rho}, \hspace{0.3cm}
	\tilde {\mathfrak a}^{7,8} \sim e^{\pm \sqrt{4 - m^2/X^2} \, \rho} ,
\eeq
where the + (-) sign corresponds to dominant (subdominant) modes, and we have disregarded terms
 polynomial in $\rho$ multiplying the exponentials.
 From these expressions one can see that the $2$-point functions will contain terms proportional to
 $\sqrt{\ell^2 - m^2/X^2}$ for $\ell=1,2,3$.

For the KS background, our calculation differs from the literature by the fact that we retained all eight sigma-model scalars,
and hence we have one additional tower of states compared to~\cite{BHM2}.
We show our results in Table~\ref{Fig:KS}, where we compare to the results quoted by~\cite{BHM2},
having normalized the states so that the comparison can be done.


\end{document}